\documentclass[10pt,journal,compsoc]{IEEEtran}
\usepackage{cite}
\usepackage{multirow}
\usepackage{graphicx}
\usepackage{color}
\usepackage{dcolumn}
\usepackage{amsmath}
\usepackage{kantlipsum}
\usepackage{url}
\usepackage[citecolor=red,linkcolor=blue,colorlinks=true,pdftex,plainpages=false]{hyperref}\usepackage{microtype}
\usepackage{verbatim}
\usepackage{colortbl}
\usepackage{amssymb}
\usepackage{arydshln}
\usepackage{multirow}
\usepackage[table]{xcolor}
\newcolumntype{K}[1]{>{\centering\arraybackslash}p{#1}}
\usepackage{booktabs}
\usepackage[flushleft]{threeparttable}
\usepackage{hhline}
\usepackage{wrapfig}
\usepackage{algorithm}
\usepackage{algorithmic}
\usepackage{mathtools}
\usepackage{placeins}
\usepackage{graphicx}
\usepackage{hyperref}
\usepackage{kantlipsum}
\usepackage{dsfont}
\usepackage[export]{adjustbox}
\usepackage{wrapfig}
\usepackage{amsmath}
\usepackage{tabularx}

\newcommand{\RNum}[1]{\uppercase\expandafter{\romannumeral #1\relax}}
\ifCLASSOPTIONcompsoc
\usepackage[caption=false, font=normalsize, labelfont=sf, textfont=sf]{subfig}
\else
\usepackage[caption=false, font=footnotesize]{subfig}
\usepackage{subcaption}
\usepackage{caption}
\usepackage{subfig}
\usepackage[utf8]{inputenc}
\usepackage{bbm}
\usepackage{tikz}
\fi
\usepackage{tikz} 
\usetikzlibrary{fit}
\usetikzlibrary{shapes}
\newcommand{\enumber}[1]{%
    \begin{tikzpicture}[remember picture]
        \node[inner sep=0pt](a){#1};
    \end{tikzpicture}%
    \begin{tikzpicture}[overlay, remember picture]
        \node[draw, red, fit=(a), ellipse, inner sep=1pt, line width=1.5pt]{};
    \end{tikzpicture}%
    }
\newcolumntype{L}{>{\centering\arraybackslash}m{1.2cm}}

\begin{document}
\title{ F\textbf{a}st De\textbf{s}ign \textbf{S}pace \textbf{E}xploration of \textbf{N}onlinear
Sys\textbf{t}ems: Part \RNum{2}}
\author{Prerit~Terway,~Kenza~Hamidouche,~and~Niraj~K.~Jha,~\IEEEmembership{Fellow,~IEEE}
\thanks{This work was done in part  under  contract from the Defense Advanced Research Projects Agency (DARPA) and Air Force Research Laboratory (AFRL). The views, opinions and/or findings expressed are those of the authors and should not be interpreted as representing the official views or policies of the Department of Defense or the U.S. Government. 
Prerit Terway and Niraj K. Jha are with the Department of Electrical Engineering, and 
Kenza Hamidouche with the Department of Operations Research and Financial Engineering, Princeton 
University, Princeton, NJ, 08544 USA, e-mail:\{pterway, jha, kenzah\}@princeton.edu.}}

\vspace{-0cm}
\IEEEtitleabstractindextext{%
\begin{abstract}
Nonlinear system design is often a multi-objective optimization problem involving search for a 
design that satisfies a number of predefined constraints.  The design space is typically very 
large since it includes all possible system architectures with different combinations of components 
composing each architecture. In this article, we address nonlinear system design space exploration 
through a two-step approach encapsulated in a framework called F\textit{a}st De\textit{s}ign 
\textit{S}pace \textit{E}xploration of \textit{N}onlinear Sys\textit{t}ems (ASSENT). In the first 
step, we use a genetic algorithm to search for system architectures that allow discrete choices for 
component values or else only component values for a fixed architecture. This step yields a coarse 
design since the system may or may not meet the target specifications. In the second step, we use an 
inverse design to search over a continuous space and fine-tune the component values with the goal of 
improving the value of the objective function. We use a neural network to model the system response. 
The neural network is converted into a mixed-integer linear program for active learning to sample 
component values efficiently. We illustrate the efficacy of ASSENT on problems ranging from
nonlinear system design to design of electrical circuits. Experimental results show that ASSENT 
achieves the same or 
better value of the objective function compared to various other optimization techniques 
for nonlinear system design by up to 53$\%$.  We improve sample efficiency by 6-12$\times$ compared to 
reinforcement learning based synthesis of electrical circuits. 
\end{abstract}

}
\maketitle

\IEEEdisplaynontitleabstractindextext

\IEEEpeerreviewmaketitle


\vspace{-.4cm}
\section{Introduction}
\label{sect:introduction}

Nonlinear system design forms the core of various applications that include healthcare, 
smart grid, transportation, and smart home \cite{sampigethaya2013aviation, 
akmandor2017keep}. Design of such systems involves solving a combinatorial optimization
problem with a large design space of various system architectures and constituent components.
It often requires solving a constrained multi-objective optimization (MOO) problem that arises in 
several disciplines like circuit design and cyber-physical system synthesis. Typically, the designer 
has access to a black-box simulator but without knowledge of the underlying dynamics governing 
system response to inputs. To reduce the development time, the designer needs to minimize the number 
of calls to the simulator that may take anywhere from minutes to hours to even days to obtain the 
response for a given choice of input. Development time can be reduced through \textit{active 
learning} of best inputs to simulate.
The input sample can be selected based on past system response to inputs and the corresponding 
outputs.

Existing techniques for nonlinear system design include reinforcement learning (RL), Bayesian 
optimization (BO), and evolutionary algorithms (EAs) like genetic algorithm (GA). However, using RL 
and BO requires formulating MOO as a weighted sum of objectives. Setting the weights requires domain 
expertise and manual tuning. Moreover, deep RL requires access to a large number of graphical 
processing units (GPUs), thus increasing development cost. BO is generally very slow as the 
complexity of generating candidate solutions increases with an increase in the number of simulations.
EAs like 
GA have been very successful in solving MOO problems. However, due to GA's inherent randomness, the 
algorithm needs to be executed over a large number of generations to meet system
specifications. Thus, they are not sample-efficient.
Surrogate-assisted EA can help improve the sample efficiency. The global search technique proposed in \cite{4033013} uses Gaussian process based surrogate model and computes probability of improvement to determine the candidate solution. This limits the methodology in \cite{4033013} to a single objective or a weighted sum of objectives. 

We propose ASSENT, a two-step methodology that builds upon CNMA\footnote{CNMA stands for Constrained 
optimization with Neural networks, MILP and Active learning} \cite{CNMA2}: a sample-efficient
exploration method that is described in detail in a sister Part I of this article. In contrast to 
CNMA that uses multiple solvers to escape local minima, in Step 1 of ASSENT, we use GA to account for 
system response over the entire design space. 
Step 1 either explores system architectures and constituent component values or only component 
values for a fixed system architecture.  We harness the benefits of GA to perform exploration in a 
discrete design space and the ease with which it can encode system architectures.  We terminate 
GA after some stopping criteria are met even if the design requirements are not yet met since GA 
is sample-inefficient and requires a large number of iterations to meet the requirements. We thus 
obtain a \textit{coarse} design at the end of the first step.

In Step 2, based on CNMA, we \textit{fine-tune} the coarse design to search for component 
values in a continuous design space.  This step uses a neural network (NN) as a surrogate function to 
model system response.  We convert the NN into a mixed-integer linear program (MILP) to incorporate 
the constraints imposed by the permissible range of inputs (i.e., component values), outputs (i.e., 
desired response), and the MILP formulation of the NN. We use this formulation to obtain an 
\emph{inverse design} of the system and find an input that satisfies the set of system constraints.  
A feasible solution of the MILP yields the input that is used to simulate the system. In the case 
of an infeasible solution, we use a random sample for simulation. The NN thus enables \textit{active 
learning} by generating high-quality input samples until system requirements are met or the sampling 
budget is exhausted.

We summarize the major contributions of this article as follows:
\begin{itemize}
\item We formulate system design as an MOO problem and propose ASSENT, a two-step method for 
sample-efficient system synthesis.
\item In Step 1 of ASSENT, we explore a large discrete design space using GA and synthesize 
a coarse design. Exploring the entire search spaces helps avoid local minima. We terminate GA even 
if the design requirements are not met. Early termination helps avoid costly simulations. 
\item In Step 2 of ASSENT, we explore a continuous design space of component values to improve the value of the objective function. We use the saved simulations from Step 1 to 
 incorporate knowledge over the entire design space. We use an NN as a surrogate to model the system response.
We use an MILP formulation of the NN to obtain an \emph{inverse 
design} of the system. The MILP formulation builds on top of CNMA where we dynamically replace the design around which we search for possible solutions. 
\item We demonstrate the efficacy of ASSENT on a wide array of design problems, ranging from 
nonlinear system design to circuit design.  System design problems include Lunar lander 
\cite{1606.01540}, Acrobot \cite{1606.01540}, 
Polak3 \cite{70250}, Waveglider \cite{kraus2012wave}, Rover path planning \cite{wang2018batched}, 
and sensor placement for a power grid \cite{openDSS}. We show that ASSENT improves the 
value of the objective function in comparison to CNMA based system synthesis by up to 54$\%$.  We also 
synthesize two-stage and three-stage transimpedance amplifiers and show that ASSENT attains a better 
value of the objective function compared to RL, BO, and human designs proposed in 
\cite{wang2018learning}.
\end{itemize}

The rest of the article is organized as follows. In Section \ref{sect:related work}, we 
discuss related work. Section \ref{sect:background} provides the necessary background, followed by 
a simple motivational example in Section \ref{sec:motivation}. In Section \ref{sect:methodology}, we 
introduce the methodology for sample-efficient nonlinear system design. We apply the methodology to 
the synthesis of nonlinear systems in Section \ref{sect:results}. 
Finally, Section \ref{sec:conclusion} concludes the article. 

\vspace{-.4cm}
\section{Related work}
\label{sect:related work}
In this section, we review previous works on solving MOO problems. We discuss weighted sum based 
techniques like RL and BO, and then MOO using EAs. We also discuss the application of 
optimization techniques to system synthesis. 
\vspace{-.3cm}
\subsection{Optimization techniques}
\label{subsec:Optimization techniques}

The most commonly employed technique to solve MOO is formulating the optimization problem as a 
weighted sum of objectives. The weights are determined using domain expertise or through 
trial-and-error until given requirements are met. On the other hand, EAs like GA do not assume weights 
are provided for different objectives but yield a Pareto-optimal front that consists of
solutions that trade off various objectives against each other.

\subsubsection{Weighted sum optimization}
\label{weighted optimization}
BO uses gradient-free search to solve optimization problems based on an acquisition function 
that enables active learning. The acquisition function points to an input that needs to be 
simulated to obtain the corresponding system output. In turn, the input/output pair is used to update 
the acquisition function.  Lyu et al.~\cite{lyu2018batch} exploit a disagreement among different 
acquisition functions to determine the next set of input points to simulate.  Upper confidence 
bound \cite{powell2019unified} is another technique that is used to obtain the best trade-off 
between exploitation and exploration when determining the next input for simulation. 

Recent works use deep RL for solving MOO in a sample-efficient manner \cite{wang2018learning}. RL requires a definition
of the reward function as a weighted sum of different objectives. A policy chooses an 
input sample for simulation based on past inputs and corresponding outputs. The recent interest
spurt in using RL combined with deep learning is due to the approach presented in 
\cite{mnih2013playing}. It is based on training a convolutional NN to learn a policy for playing Atari 
games. It was extended to the continuous action space in \cite{lillicrap2015continuous} to solve 
simulated physics tasks and learn end-to-end policies that are more sample-efficient than those 
discussed in \cite{mnih2013playing}.

\subsubsection{MOO}
\label{weight free optimization}

EA is a widely used gradient-free search technique for obtaining Pareto-optimal solutions to an MOO 
problem. GA, a type of EA, is known to be suitable for such problems \cite{zebulum2018evolutionary}. 
A seminal work in this field is the NSGA-\RNum{2} algorithm \cite{996017} that uses a fast 
nondominated sorting approach to reduce computational complexity by orders of magnitude. EAs have 
several variants, e.g., differential evolution that perturbs randomly selected population members 
based on the difference between selected individuals \cite{kenneth1999price}, swarm intelligence that 
includes ant colony optimization \cite{4129846}, and particle swarm optimization 
\cite{kennedy1995particle}. The sampling efficiency of GA is improved using surrogate models in \cite{4033013}.
Promising individuals are filtered using a Gaussian process based surrogate model in the first level. The second level uses radial basis function for gradient based search. 

\vspace{-.3cm}
\subsection{System synthesis}
\label{subsec:System synthesis}

EA was widely used in the 1990's and early 2000's for the synthesis of electrical circuits 
\cite{zebulum2018evolutionary}.  The use of EA to obtain the topology of analog circuits was
pioneered in \cite{koza1997automated}.  Parallel GA and circuit-construction primitives 
were used to create circuit graphs to evolve designs for an analog filter and amplifier
in \cite{lohn1999circuit}. 
Synthesis of amplifiers and filters was
done by solving constrained MOO with NSGA-\RNum{2} in \cite{nicosia2007evolutionary}. 

More recent works have primarily focused on determining component values for a fixed architecture. 
The method proposed in \cite{wang2018learning} uses deep deterministic 
policy gradient (DDPG), a form of RL, for sample-efficient synthesis of two-stage and three-stage 
transimpedance amplifiers.  The method in \cite{settaluri2020autockt} 
proposes deep RL combined with transfer learning over a sparse design space to synthesize analog 
circuits. The method in \cite{lyu2018batch} uses BO with an ensemble of acquisition functions to 
tackle complex mathematical functions and synthesize electrical circuits.  We instead rely on 
CNMA \cite{CNMA2} that uses an MILP formulation of the NN for selecting component values to synthesize 
systems.

The drawbacks of existing search techniques are as follows. (1) The system design problem is 
formulated as a weighted sum of multiple objectives. In the
optimization process, the weights are determined using domain expertise or through 
simulations, making the design procedure inefficient. (2) Most system design formulations assume that 
the system architecture is fixed and only focus on selecting component values, thereby limiting the 
design space and possibly missing out on novel designs. (3) GA is sample-inefficient and often needs 
to repeat costly simulations multiple times to obtain an acceptable design. (4) BO based design 
techniques require a large amount of time to select the next sampling point for simulation 
\cite{wang2018learning}, thereby making the optimization process very slow.

\vspace{-.4cm}
\section{Background}
\label{sect:background}

In this section, we discuss preliminary material that our work builds on. We discuss the formulation 
of system design as an MOO problem. Next, we discuss GA and its applicability to system architecture 
search. Finally, we give a brief review of CNMA for inverse design of a system. The
sister article \cite{CNMA2} describes CNMA in great detail.

\vspace{-.3cm}
\subsection{System design through multi-objective optimization}
\label{multi-objective}
When designing a system, the designer is interested in achieving the desired objectives subject to 
given constraints.  MOO is one of the ways to formulate this problem.  The MOO solutions 
enable best trade-offs among competing objectives. They constitute a \emph{nondominated} set and lie 
on a surface called the \emph{Pareto front} \cite{van1998evolutionary}.  More formally, the  system 
design problem is formulated as follows:

\begin{equation}
\label{eq: MultiObjEquGeneric}
    \begin{aligned}
    & \underset{\boldsymbol{x}, \boldsymbol{y}}{\text{minimize}}
    & & f_{m}(\boldsymbol{x}, \boldsymbol{y}), \ m=1,2, \ldots, M \\
    & \text{subject to}
    & & g_{j}(\boldsymbol{x}, \boldsymbol{y}) \geq 0, j=1,2, \ldots, J \\
    &&& h_{k}(\boldsymbol{x}, \boldsymbol{y})=0, k=1,2, \ldots, K \\
    &&& x_{i}^{L} \leq x_{i} \leq x_{i}^{U}, i=1,2, \ldots, n
    \end{aligned}
\end{equation}
\noindent
where $\boldsymbol{y}$ represents the design space over all possible architectures or the ways 
components can be connected to form a system, $\boldsymbol{x}$ is the design space over all possible 
component values, and $f_{m}(\boldsymbol{x}, \boldsymbol{y})$ is the $m^{th}$ objective.  The system 
must also satisfy $J$ inequality constraints given by $g_{j}(\boldsymbol{x}, \boldsymbol{y})$ and 
$K$ equality constraints given by $h_{k}(\boldsymbol{x}, \boldsymbol{y})$. For instance, when 
designing a Lunar lander \cite{1606.01540}, the goal is to maximize the reward determined by the 
position at which the module lands subject to constraints on fuel consumption and time taken to land. 
The range of values of available components constrains the inputs. 
The lower bound for value $x_{i}$ of component $i$ is labeled as $x_{i}^{L}$ and 
its upper bound as $x_{i}^{U}$.  

\vspace{-.3cm}
\subsection{Genetic algorithm}
\label{GA}

System designs often rely on human expertise.  However, this may lead to failure to find a design 
that meets all system requirements or require long simulation runs to explore the large design 
space.  In the first step of ASSENT, we harness GA's exploration capability and ease of encoding an 
architecture~\cite{goldberg1988genetic} to solve an MOO formulation of system design.
GA evolves a population of individual solutions through multiple generations.  The main steps of GA 
are as follows. (1) Each individual in the first generation of a population is represented with a 
\textit{chromosome}. A chromosome is a sequence of genes. (2) Each individual is evaluated based on how well it meets its objectives and constraints. (3) A subset of these individuals is selected to produce children for the next generation. (4) Pairs of randomly chosen individuals (parents) from this subset undergo
reproduction using \emph{crossover} by combining the genes of the parents to create children. (5) The genes of each child are \emph{mutated} by perturbing them with low probability to 
facilitate exploration across various dimensions. (6) The best performing individuals in the current generation and the children (after
Step 5) are retained. (7) Steps 2-6 are repeated until one of the stopping criteria is met.

\begin{figure}[!ht]
    \includegraphics[scale=0.25]{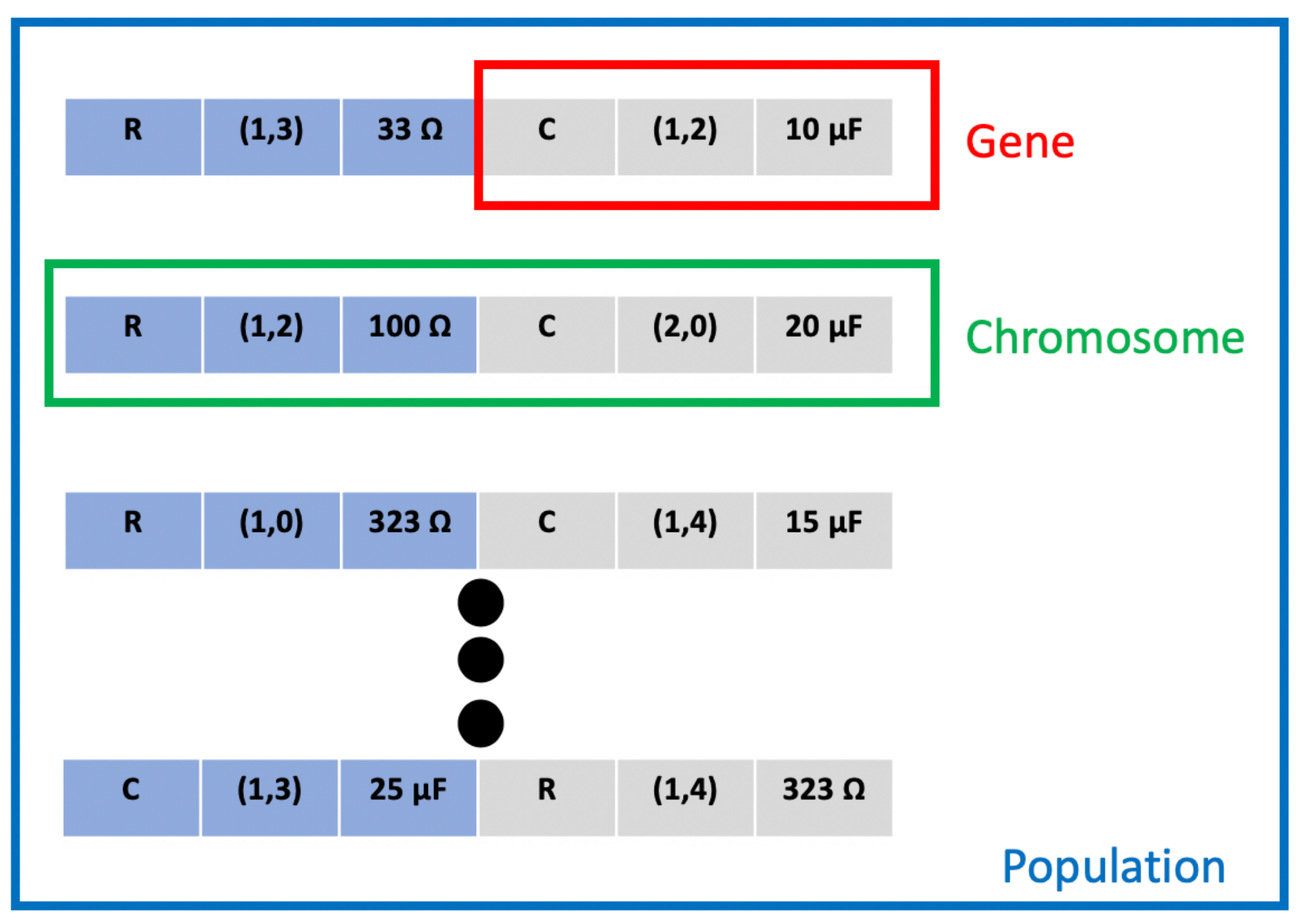}
    \centering
    \caption{Details of different elements of GA. Each component is color-coded 
(resistor in blue and capacitor in grey). Gene depicted in a red box, chromosome in a green box, and 
population in a blue box.}
\label{fig:GA details}
\end{figure}

\begin{figure}[!ht]
  \centering
    \includegraphics[width=0.40\textwidth]{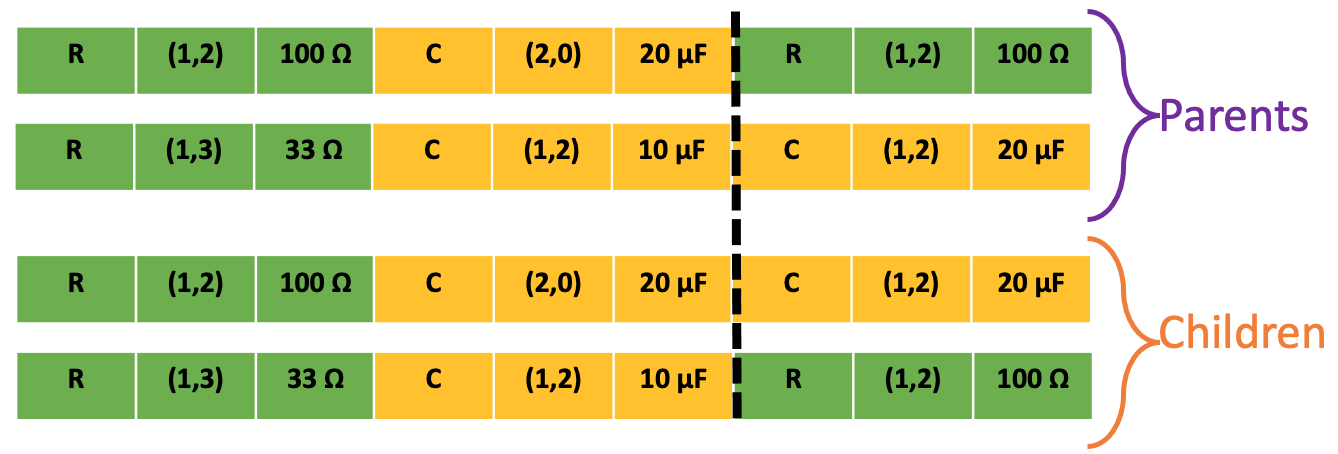}
    \caption{Crossover operation to produce children from parents. Crossover point shown with a dotted line.}
    \label{fig: cross}
\end{figure}

\begin{figure}[!ht]
  \centering
    {%
      \includegraphics[width=0.40\textwidth]{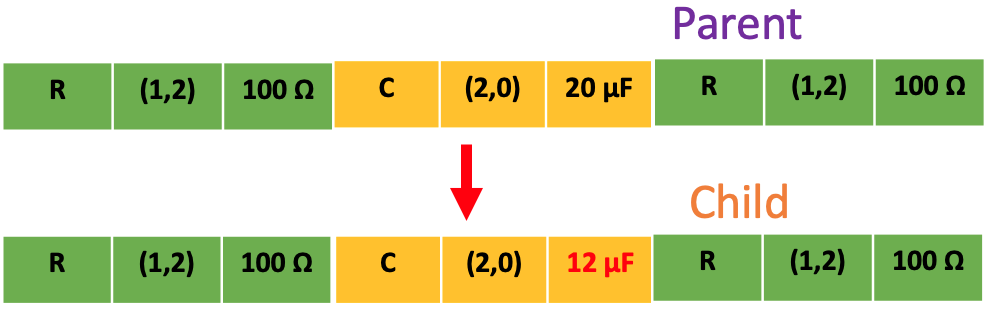}}
  \caption{Mutation of an individual. The capacitance changes from 20 $\mu$F to 12 $\mu$F. The changed value is shown in red.}
  \label{fig: mut}
\end{figure}

In ASSENT, a \emph{chromosome} corresponds to a design choice for the complete system. The design 
choice may represent the system architecture and its component values or only the  component values 
for a fixed architecture. A \emph{gene} encodes details of a particular component, like its type, 
connecting nodes, and value.  We evaluate the design through a simulator. We use tournament selection 
to select the individuals that undergo reproduction \cite{blickle1996comparison}. 
Fig.~\ref{fig:GA details} shows a population of individuals, i.e., chromosomes, and genes. Here, the
system architecture corresponds to an electrical circuit. The gene shown in the red box depicts a 
capacitor with a capacitance of 10 $\mu F$ connected between nodes 1 and 2. A sequence of genes forms 
a chromosome or an individual shown in the green box.  The blue box shows the entire population 
within a generation. Fig. \ref{fig: cross} shows the crossover operation. The dotted line shows the crossover point. The genes of the two parents are exchanged at the crossover point to produce children.
Fig. \ref{fig: mut} shows mutation of an individual. The capacitance changes from 20$\mu$F to 12$\mu$F. The changed value is shown in red. GA requires many simulations to obtain a design that meets all the 
specifications, thereby pointing to the need for a sample-efficient design procedure. 

\vspace{-.5cm}
\subsection{Inverse system design using NN}
\label{MILP Formulation}
We overcome the sample inefficiency of GA by terminating it at an intermediate stage (before 
necessarily reaching a valid or acceptable design) and then using a sample-efficient  
search method to meet system requirements.  Akintunde et al.~\cite{akintunde2018reachability} proposed 
an MILP formulation of an NN with ReLU activation in the context of neural agent-environment systems to solve the
problem of \emph{reachability} through an NN-based policy trained using RL.  Reachability indicates 
whether the NN can output the desired values using permissible inputs.  An NN is converted into an 
MILP by representing all hidden neurons with constraints defined as follows:

\begin{equation}
\label{eq:MILPConversion}
  \begin{aligned} C_{i}=\left\{\bar{x}_{j}^{(i)}\right.& \geq W_{j}^{(i)}
\bar{x}^{(i-1)}+b_{j}^{(i)}, \\ \bar{x}_{j}^{(i)} & \leq W_{j}^{(i)}
\bar{x}^{(i-1)}+b_{j}^{(i)}+Q \bar{\delta}_{j}^{(i)}, \\ \bar{x}_{j}^{(i)} &\left.\geq 0,
\bar{x}_{j}^{(i)} \leq Q\left(1-\bar{\delta}_{j}^{(i)}\right), j=1, \ldots , L^{(i)} \right\}. \\  \end{aligned}  
\end{equation}

\noindent
In Eq.~(\ref{eq:MILPConversion}), $\forall{i,j}$, $\bar{x}_{j}^{(i)}$ corresponds to the $j^{th}$ 
neuron in the $i^{th}$ layer, $L^{(i)}$ is the number of neurons in the $i^{th}$ layer, $W_{j}^{(i)}$ represents weights that determine the input to
$\bar{x}_{j}^{(i)}$, $\bar{x}^{(i-1)}$ represents outputs from the $(i-1)^{th}$ layer, 
${b}_{j}^{(i)}$ is the bias for neuron $\bar{x}_{j}^{(i)}$, $Q$ is larger than the largest possible 
magnitude of $W_{j}^{(i)} \bar{x}^{(i-1)}+b_{j}^{(i)}$, and $\bar{\delta}_{j}^{(i)}$ is defined as 
follows:
\begin{equation}
\label{eq:delta}
    \begin{aligned}
   \bar{\delta}_{j}^{(i)} \triangleq\left\{\begin{array}{l}0 \text { if } \bar{x}_{j}^{(i)}>0 \\ 1 \text { otherwise }\end{array}\right. 
   \end{aligned} 
\end{equation}
 The constraints imposed by the hidden neurons of the network are
obtained from the union of all the constraints ($C_{i}$) shown in Eq.~(\ref{eq:MILPConversion}).

CNMA \cite{CNMA2} uses the above formulation to determine the component values of a given system
architecture with the aim of achieving the best value of the objective function.  The designer specifies the objective and constraints on the inputs and the desired output.
A feasible solution to the MILP determines the component values required to perform the 
simulation. In the case of an infeasible solution, the system is simulated using a random sample. 
Next, we illustrate ASSENT using a motivational example that involves designing a low-pass filter.

\vspace{-.3cm}
\section{Motivation}
\label{sec:motivation}
We propose ASSENT as a solution to the nonlinear system design problem. We illustrate its working 
through the design of a low-pass filter.
The objective is to obtain a \textit{unity gain ($0\  \rm dB$)} and a \textit{bandwidth of $1~\rm kHz$} 
while \textit{minimizing} the number of circuit components. We first synthesize a coarse design in Step 1 that is fine-tuned in Step 2.
\vspace{-.4cm}
\subsection{Coarse design}
\begin{figure}[!ht]
    \includegraphics[scale=0.11]{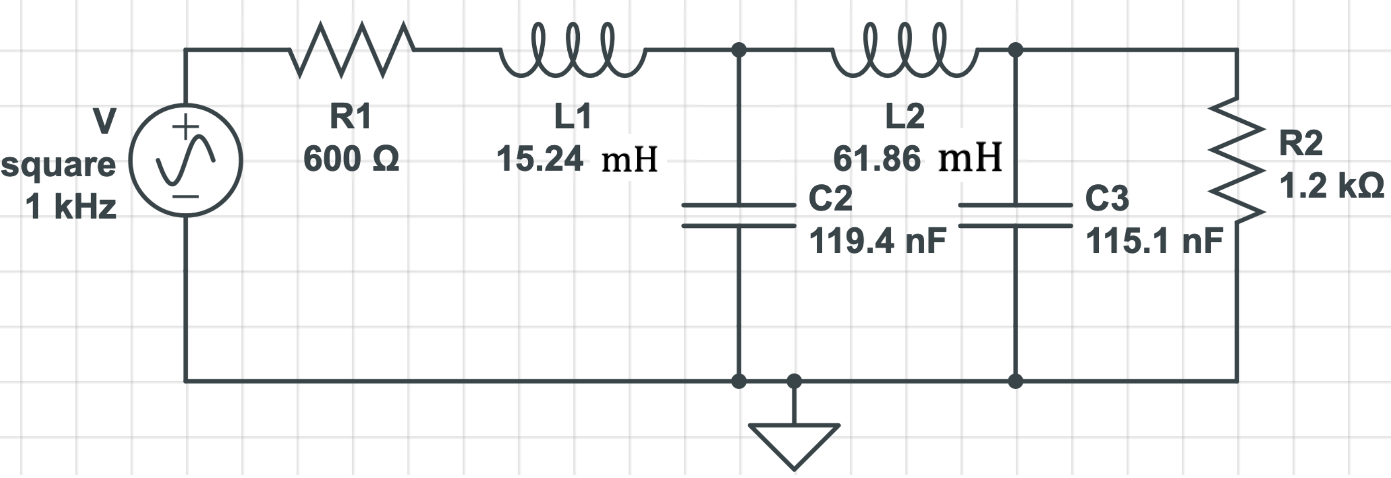}
    \centering
    \caption{Low-pass Butterworth filter architecture and component values used as a seed design.}
\label{fig:ButterworthFilter}
\end{figure}

We assume the following discrete components are 
available in Step 1: (1) resistors [1, 10, 600, 1200] $\Omega$, (2) capacitors [1e-12, 119.37e-9, 
155.12e-9, 1e-5] F, and (3) inductors [1e-6, 10e-3, 15.24e-3, 61.86e-3] H.
The seed design is a Butterworth low-pass filter adapted from \cite{venturini_2015} and shown in 
Fig.~\ref{fig:ButterworthFilter}. We evolve it through GA till the point it gets close to meeting
the specifications.  We use a maximum of five nodes and 10 components in the circuit
architecture to expand the design space even though we know that a low-pass filter can be
designed with just three nodes and two components (besides the supply).
The total number of choices for a gene is given by $(\# \textit{component types})\times
(\# \textit{connecting points})\times(\# \textit{different component values}) = 
3\times5^2\times4 = 300$ (each component has two connecting points resulting in $5^2$ connecting points) ~\cite{zebulum2018evolutionary}. In our example, since there may be 10 
components in the circuit, the design space size is $300^{10}$.
Hence, we need to search the large design space efficiently.

\begin{figure}[!ht]
    \includegraphics[scale=0.28]{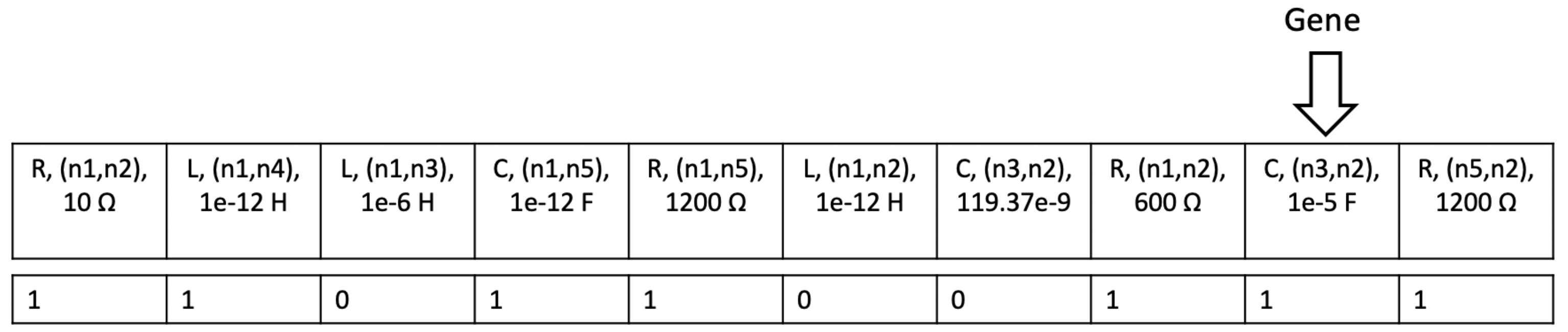}
    \centering
    \caption{Chromosome representation of a low-pass filter. The top row represents details of all 
the 10 components. It shows the component type given by (R, L, C) and its connecting nodes 
($n$1, $n$2, $n$3, \ldots) and value.  The bottom row indicates whether the component is active (1) 
or inactive (0) in the circuit.}
\label{fig:LowPassGene}
\end{figure}

Fig.~\ref{fig:LowPassGene} shows the chromosome representation of a circuit. A gene encodes details of a  component, including its type, connecting nodes, and value (shown in the top row). The 
bottom row shows whether the component is active (1) or inactive (0)~\cite{zebulum2018evolutionary}.  We use the following objectives to find the coarse design by searching the architecture and 
component space using GA. (1) Weighted sum of the difference between the desired \textit{magnitude} response of a 
first-order low-pass filter and the observed response. We set the weights in the \textit{passband} 
to $40$ and $1$ in the \textit{stopband} to give more importance to the response in 
the \textit{passband}.  Note that since GA synthesizes a coarse design, other values may work too. We use a weighted sum here to cover the entire frequency range. (2) Weighted sum of the difference between the desired \textit{phase} response of a first-order 
low-pass filter and the observed response. The weights are the same as above. (3) Number of \textit{active} components: a lower value implies fewer components.


\begin{figure*}[htbp]
\centering
\subfloat[]{\includegraphics[height=1.0in, width=2.in]{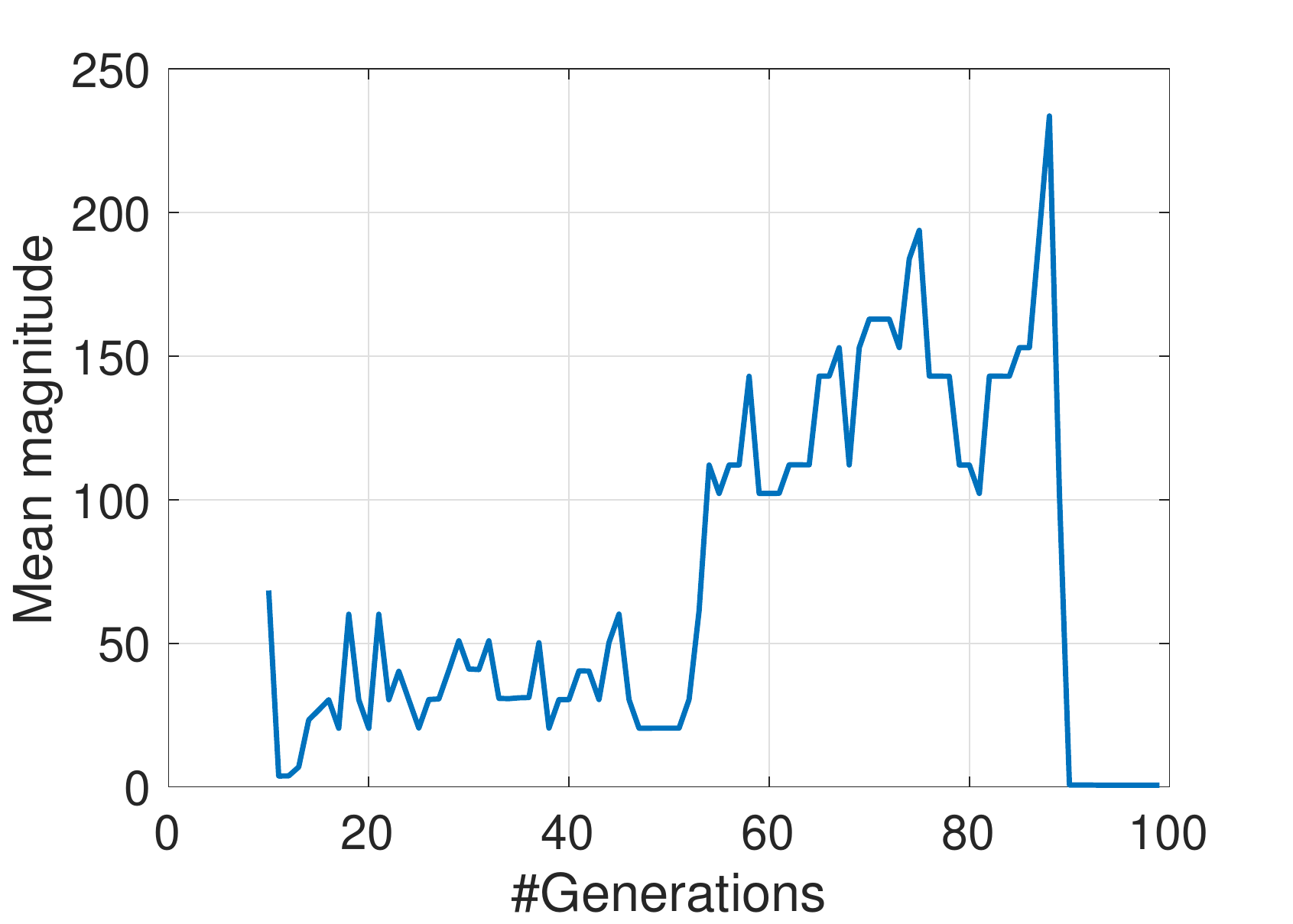}}
\subfloat[]{\includegraphics[height=1.0in, width=2.in]{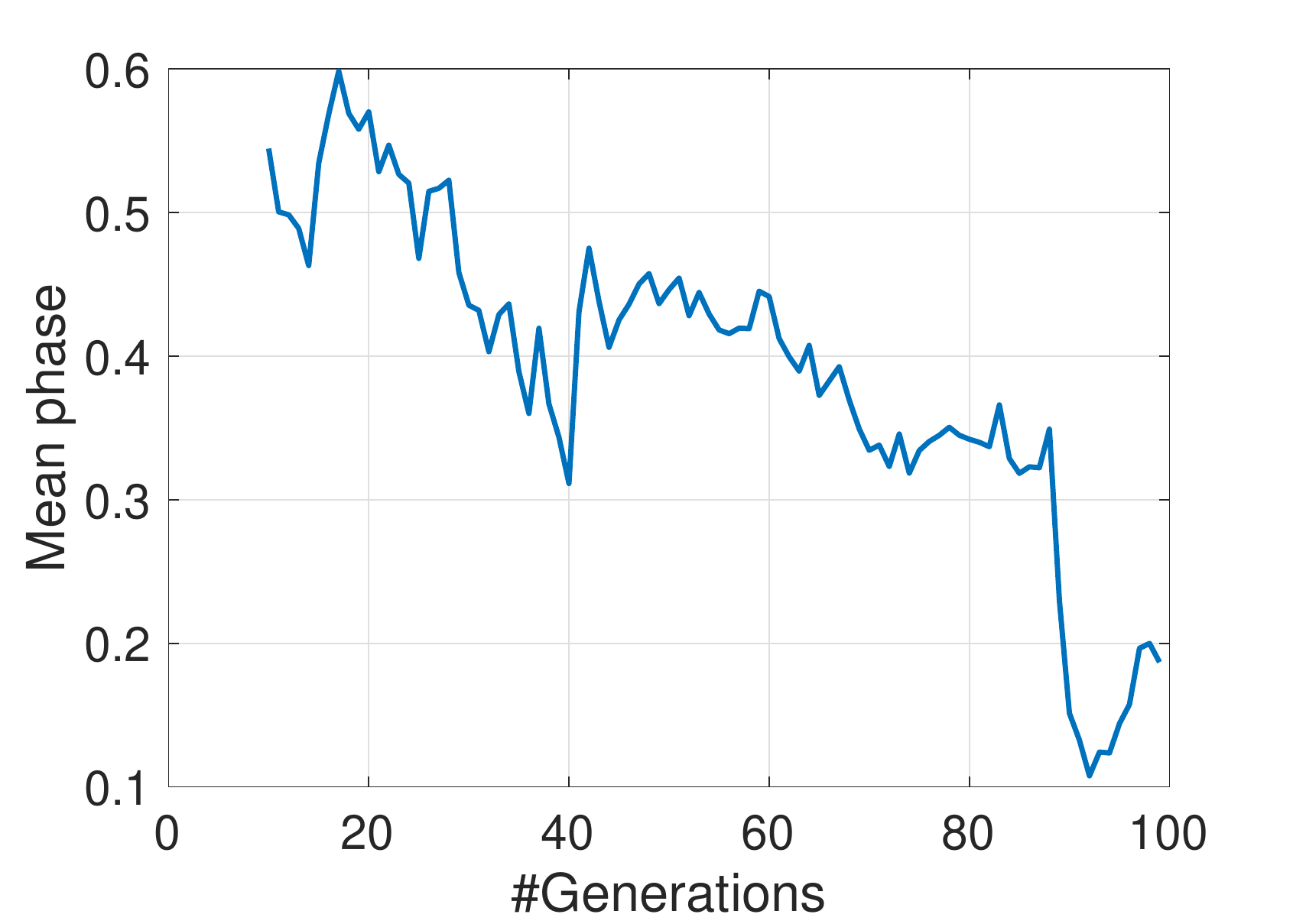}}
\subfloat[]{\includegraphics[height=1.0in, width=2.in]{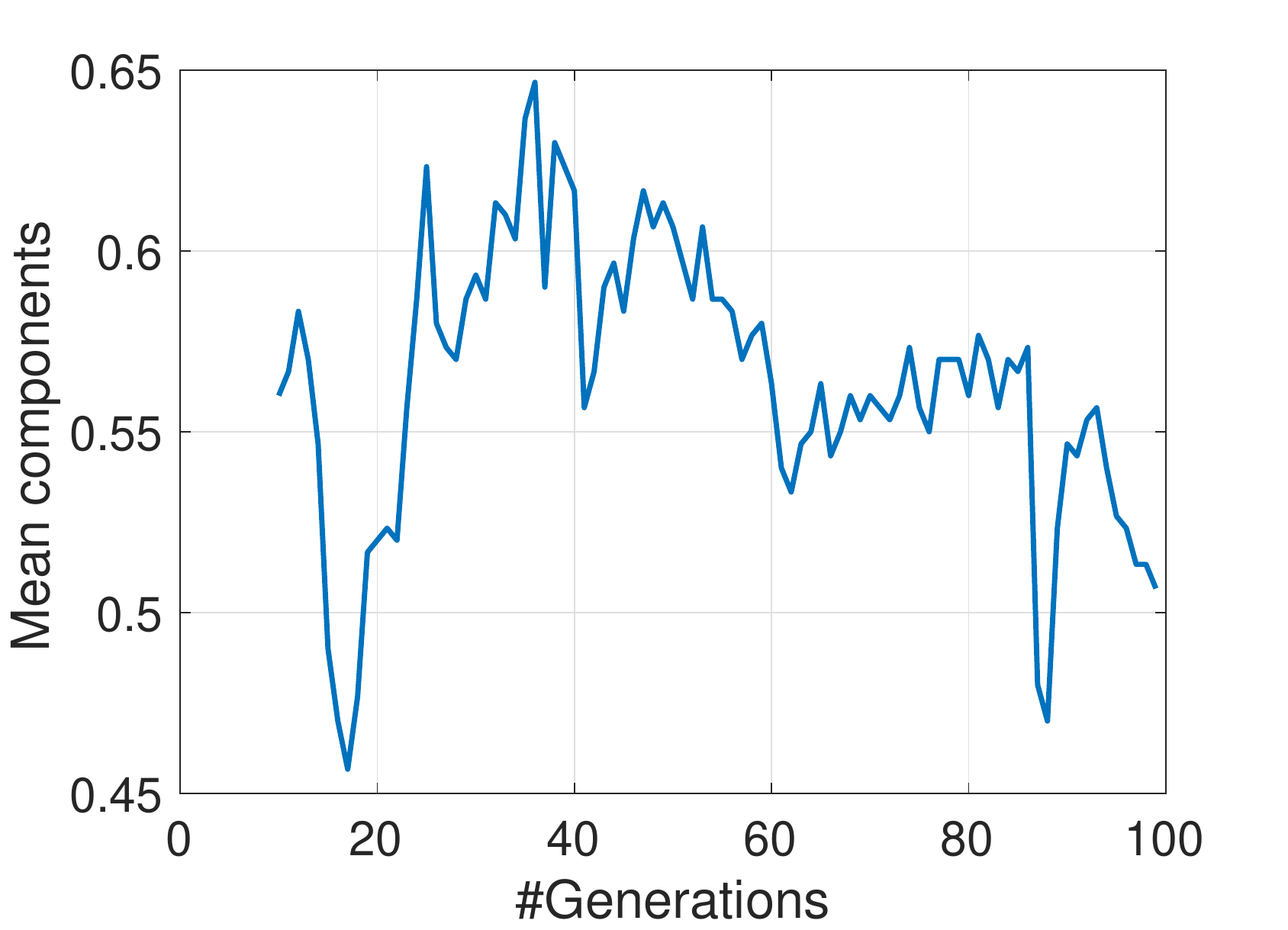}}
\caption{Scaled mean objectives across all individuals in a generation for (a) magnitude, (b)
phase, and (c) component values from the $10^{th}$ generation onwards for architecture search 
for low-pass filter design.}
\label{fig:GAEvolutionLowPass}
\end{figure*}

\begin{figure}[!ht]
    \includegraphics[scale=.50]{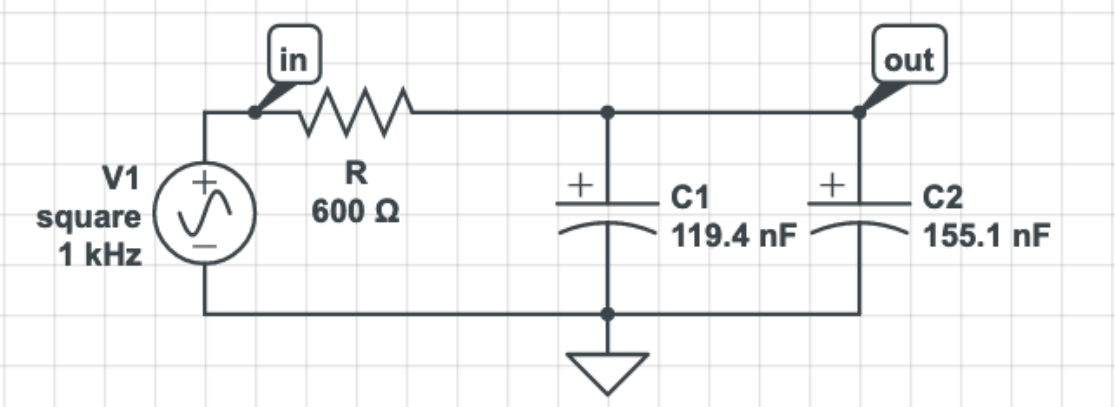}
    \centering
    \caption{Low-pass filter architecture evolved by GA.}
\label{fig:GASynthLowPass}
\end{figure}

We use GA for architecture search using 50 individuals comprising the seed design and other randomly 
generated individuals evolved over 100 generations. 
Fig.~\ref{fig:GAEvolutionLowPass} shows the scaled mean (with respect to the seed design) of 
the three objectives for the entire population from the $10^{th}$ generation onwards. We plot
the mean objective values after the $10^{th}$ generation because the initial GA generations have very 
high objective values.  The $x$-axis shows the generation number and the $y$-axis shows the mean value 
for the three objectives.  Since the mean values are scaled, the $y$-axis has no unit.     
There is a trade-off among the three objectives across generations, as evident from the figure. 
After 100 generations, we select a nondominated individual from the Pareto front based on the first 
two objectives (magnitude and phase).  This individual (circuit) has a bandwidth of 966 Hz, phase 
of -46 degree, a gain of 0 dB, and a total of three components. Since this is a coarse design, it does 
not meet the requirements yet. Fig.~\ref{fig:GASynthLowPass} shows the GA-evolved circuit. The circuit 
synthesized by GA can be simplified using human intervention by replacing the two parallel capacitors 
with a single equivalent capacitor, thus yielding a standard low-pass filter.  The next step is to 
fine-tune the component values in a continuous design  space.

\vspace{-.3cm}
\subsection{Fine-tuning}
\begin{figure}[ht]
    \includegraphics[scale=0.15]{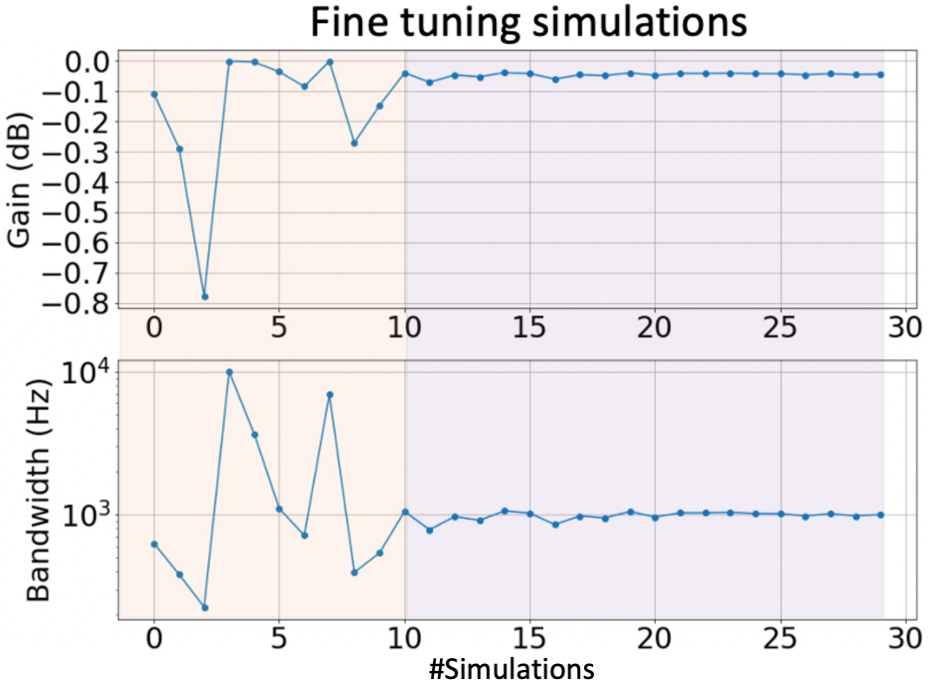}
    \centering
    \caption{Gain and bandwidth during fine-tuning using Step 2.}
\label{fig:CNMASimulation}
\end{figure}

\begin{figure}[ht]
    \includegraphics[scale=0.2]{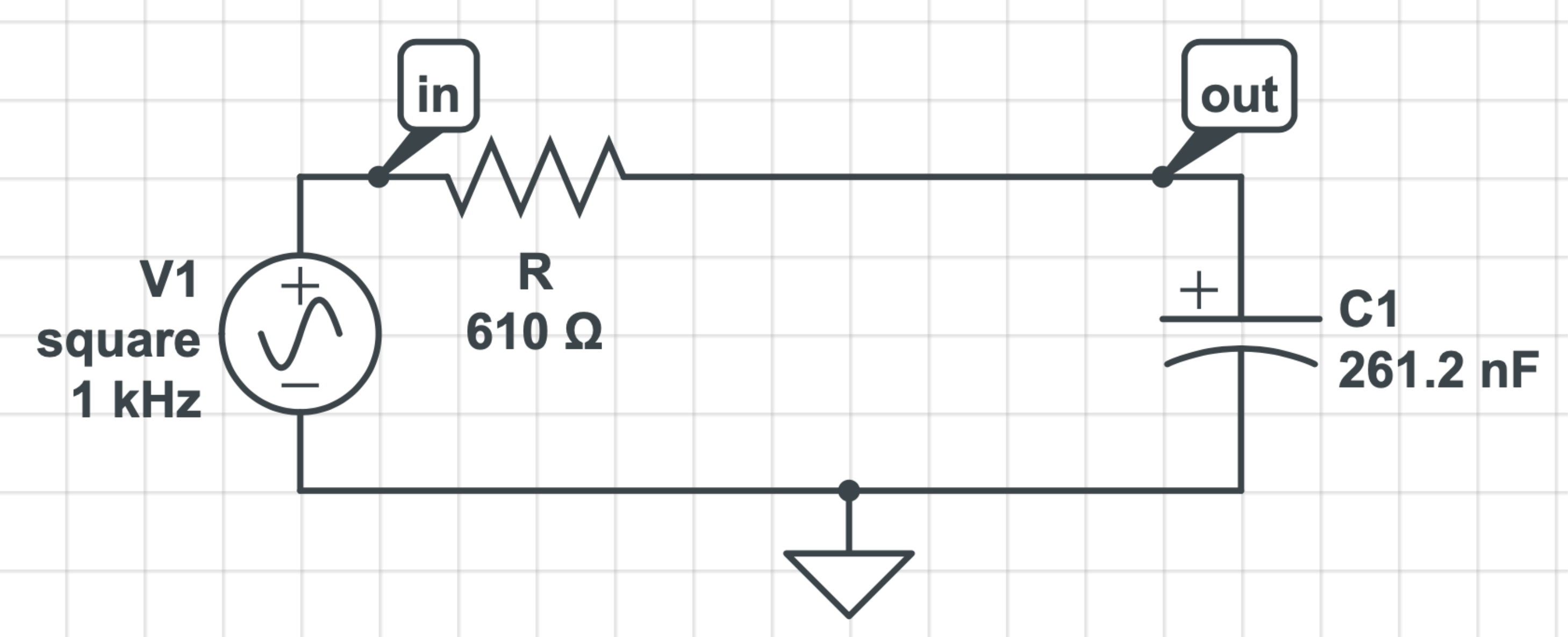}
    \centering
    \caption{Low-pass filter architecture (after human intervention on the GA schematic) at the end 
of the second step.}
\label{fig:CNMALPF}
\end{figure}

We perform fine-tuning through an NN with a hidden layer consisting of 100 neurons 
 converted into an MILP. The NN inputs are component values (resistor and capacitor) 
derived from a feasible solution to the MILP or a random sample if a feasible 
solution does not exist. The continuous design space is $[400, 800] \Omega$ for \textit{resistor}
and $[0.01, 1]\rm \mu F$ for \textit{capacitor}. Since this is a simple design, 
through prior knowledge, we know that a low-pass filter can be synthesized within these ranges of 
component values to meet the specifications. The targeted outputs of the NN are gain, bandwidth, and 
response of the filter at 200, 500, and 2000 Hz. The responses at these frequencies sufficiently 
capture the behavior of the filter below and above the desired bandwidth. The output constraints 
are as follows: gain (from input to output) in the range $[-0.92, 0.83]$ dB and bandwidth in the range 
$[990, 1010]$ Hz. We generate 10 random input samples to initialize the NN for training to 
minimize the MSE.  The solutions suggested by MILP meet the requirements after 20 more 
simulations. We show these simulations (10 for initialization of the NN and 20 during the MILP step) 
in Fig.~\ref{fig:CNMASimulation}: initialization points in the orange part and MILP ones in the blue 
part.  The outcomes during initialization are random as they are far from meeting the 
requirements. However, the points suggested by the MILP have a response closer to the requirements 
until finally, the suggested point meets the specification. Fig.~\ref{fig:CNMALPF} shows the component values on termination of Step 2.


\vspace{-.30cm}

\section{Synthesis Methodology}
\label{sect:methodology}
In this section, we describe ASSENT in detail.  The first step 
explores a large design space using GA. The outcome of the first step is a \emph{coarse} design that 
is fine-tuned in Step 2 using a sample-efficient search to obtain component values that satisfy all 
the constraints and try to improve the value of the objective function attained in Step 1.

\vspace{-.2cm}
\subsection{Step 1: Coarse design}
\label{sec: coarseDesign}
Design of systems with multiple choices for architectures and component values requires search in 
a large design space.  We achieve this through GA mainly due to its ease of encoding architectures 
and component values. GA also helps in exploring the entire design space, thus avoiding local minima. We discretize the large design  space to obtain a coarse design.
The coarse design may even violate hard constraints or attain an inferior value of the objective 
function compared to other synthesis methodologies. 

\begin{figure}[ht]
    \includegraphics[scale=.38]{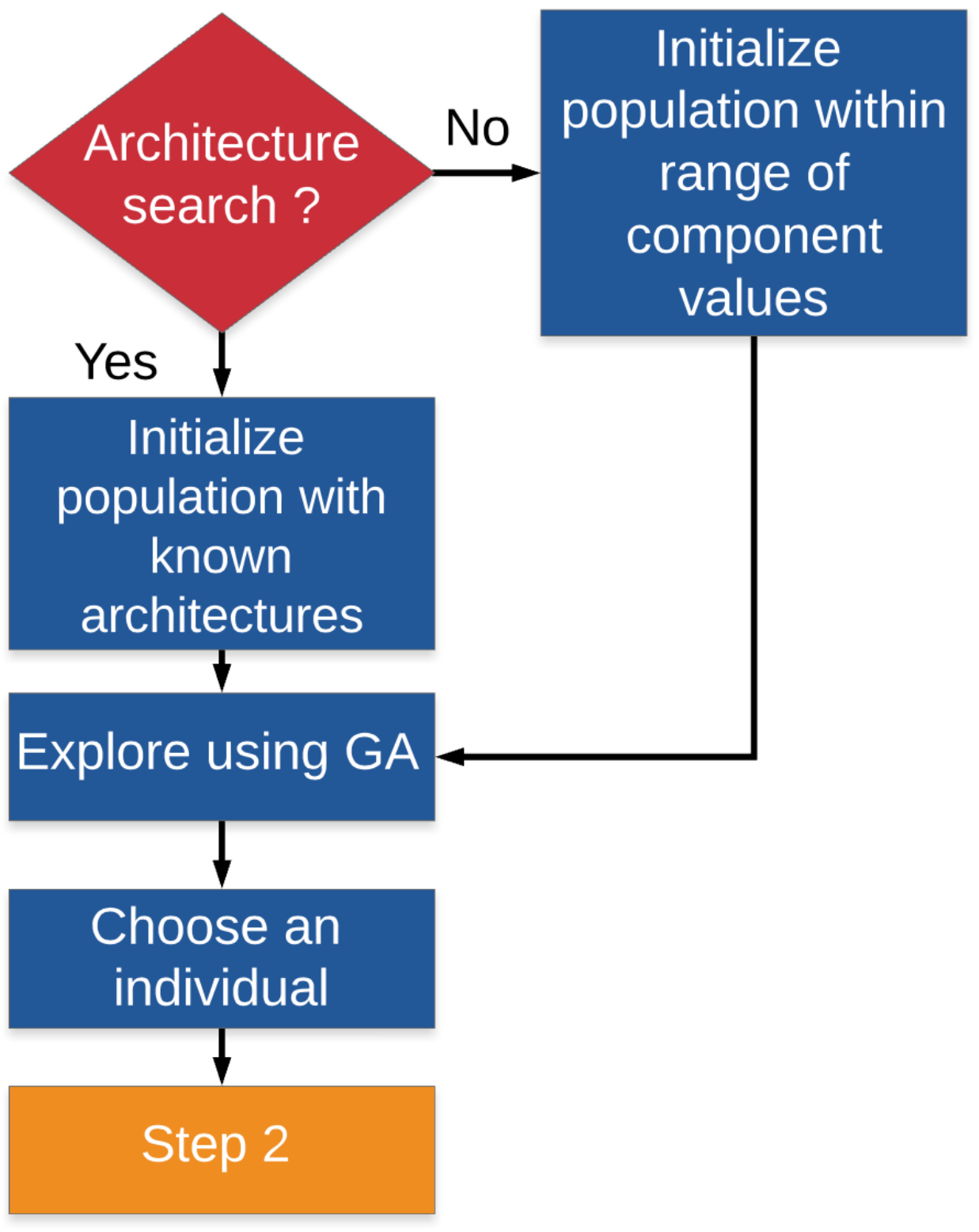}
    \centering
    \caption{Evolution using GA.}
\label{fig:Step 1 GA}
\end{figure}

We show the flow involved in the synthesis of the coarse design in Fig.~\ref{fig:Step 1 GA}. We 
evolve individuals across generations using NSGA-\RNum{2} \cite{996017} that yields nondominated
solutions of the system design formulated as an MOO problem.  In architecture search, we initialize 
some individuals in the first 
GA generation with seed design(s) from the literature to exploit prior knowledge. When performing only 
component selection, we initialize the individuals in the first generation with component values to 
cover the design space.  This enables comparisons with other designs from the literature that do not 
use prior knowledge.

\vspace{-.0cm}

During \emph{{architecture search}}, a \emph{gene} has three constituents: type of component, 
nodes connecting the component, and the component value, as shown in Fig.~\ref{fig:LowPassGene}. In 
the case of \emph{{component selection}}, the gene encodes the value of the component in the 
\emph{chromosome} that represents the circuit. 

Algorithm \ref{alg:GAArchAlgorithm} describes architecture search  using GA. We 
generate \textit{P} individuals in a generation. Some of these are from seed designs (\textit{S})
and others randomly generated.  We generate random individuals by selecting a random 
component, choosing the number of \textit{nodes} (e.g., a resistor requires two nodes whereas a MOSFET 
requires three nodes) connected to that component, and the component value. We select the component 
values for random individuals from quasi-random \textit{Sobol} samples in the range (\textit{R}) of 
each component. Sobol samples are distributed uniformly over a unit hypercube 
\cite{burhenne2011sampling}.  Then we scale the samples to lie within the specified range for each 
component. We post-process individuals to ensure that some components connect to the fixed 
terminals of the architecture, e.g., \emph {ground, supply voltage, input/output terminals in
the case of a circuit}, to ensure a valid design. We simulate all individuals in a generation to 
compute the objective functions. If the simulation is expensive, we utilize multiple cores and run GA 
in parallel. We store all the simulations in a buffer \emph{B} to be used in Step 2, if necessary. We also use the buffer as a lookup table in case the same design needs to be simulated again in Step 1.
Next, we rank the individuals using the \textit{NSGA-\RNum{2}} 
algorithm \cite{996017}. We use tournament selection for selecting some individuals in a generation 
to undergo reproduction through \emph{crossover} with a probability of \emph{cross} and 
\emph{mutation} with a probability of \emph{mut} to produce \textit{P} children.
We use \textit{NSGA-\RNum{2}} again to select \textit{P} individuals out of the \textit{2P} 
individuals for the next generation. This process continues until one of the stopping criteria
(\textit{stop}) is met. These criteria are based on individuals not improving over a fixed number 
of generations, exhausting the simulation budget, or on attaining the required performance. Finally, 
we select one individual from the final generation based on a performance metric. The metric 
(lower the better) in our case is one of the multiple  objectives or constraints. We select an 
individual from the last generation since ranking using \textit{NSGA-\RNum{2}} ensures that we never 
lose an individual from the Pareto front within a generation. We select only one individual since GA 
only synthesizes a coarse system design, although it is also possible to select another individual to 
undergo fine-tuning in the next step. 

\begin{algorithm}[h]
    \caption{Step 1: Architecture synthesis using GA}
    \label{alg:GAArchAlgorithm}
    \begin{algorithmic}[l]
        \REQUIRE \textit{S}: Seed design(s); \textit{N}: $\#$generations; \textit{P}: population size; \textit{C}: Components; \textit{R}: Component range; \textit{nodes}: Nodes; \textit{max\_comp}: Max $\#$ components; \textit{stop}: Stopping criteria; \textit{B}: buffer to store simulations; \textit{mut}: mutation probability; \textit{cross}: crossover probability
        \STATE - Generate \textit{Sobol} samples for \textit{C} within \textit{R}
        \STATE - Initialize individuals with \textit{S} and others randomly
        \WHILE{not \textit{stop}} 
            \STATE - Compute \textit{objectives} for all \textit{P}
            \STATE - Store input and the corresponding output in \textit{B}
            \STATE - Rank \textit{P} using \emph{NSGA-\RNum{2}}
            \STATE - Use tournament selection to create mating population of size \textit{P}
            \STATE - Use reproduction based on \emph{crossover} and \emph{mutation} to create 
\textit{P} children
            \STATE - Select \textit{P} from \textit{2P} members using \emph{NSGA-\RNum{2}}
        \ENDWHILE
        \ENSURE Best individual from the final generation using a metric, and buffer B
    \end{algorithmic}
\end{algorithm}

Component selection for a fixed architecture uses a setup similar to Algorithm 
\ref{alg:GAArchAlgorithm}. 
Instead of initializing some individuals in the first generation by seed designs, we initialize all 
the individuals using Sobol samples. The rest of the procedure remains the same. 

\vspace{-.3cm}
\subsection{Step 2: Fine-tuning}
\label{CNMA}

\begin{figure}[h]
    \includegraphics[scale=0.23]{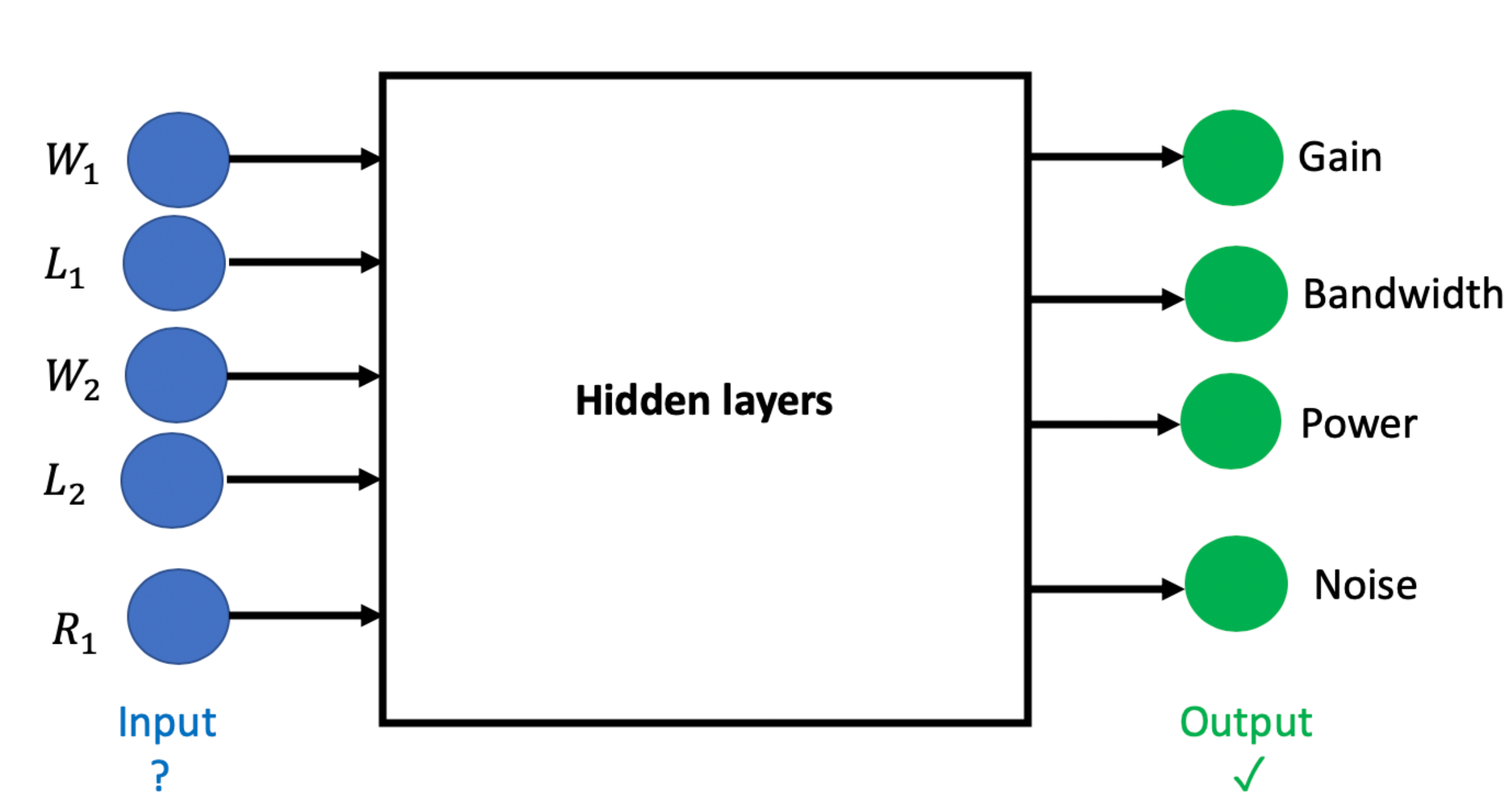}
    \centering
    \caption{Overview of the inverse design used in Step 2.} 
\label{fig:CNMAHighLevelWhy}
\end{figure}

The design obtained from Step 1 may not always meet the hard constraints. Even if it meets the hard 
constraints, it is possible to improve the value of the objective function by searching in a 
continuous design space.  When using Step 1 for architecture search, 
we use human intervention to refine the synthesized design if needed. We fine-tune this design
through a modified version of CNMA \cite{CNMA2} by searching in a continuous space. In contrast to 
CNMA where the component selection stops on exhausting the simulation budget, we adaptively replace 
the design from Step 1 to improve the value of the objective function. 
Fig.~\ref{fig:CNMAHighLevelWhy} shows a high-level overview of this step through an example. The 
designer specifies the system requirements shown on the right in green. The feasible solution of the 
MILP determines the potential component values shown on the left in blue that achieve the desired 
response.

\begin{figure}[!ht]
    \includegraphics[width=8cm, height = 6.10cm]{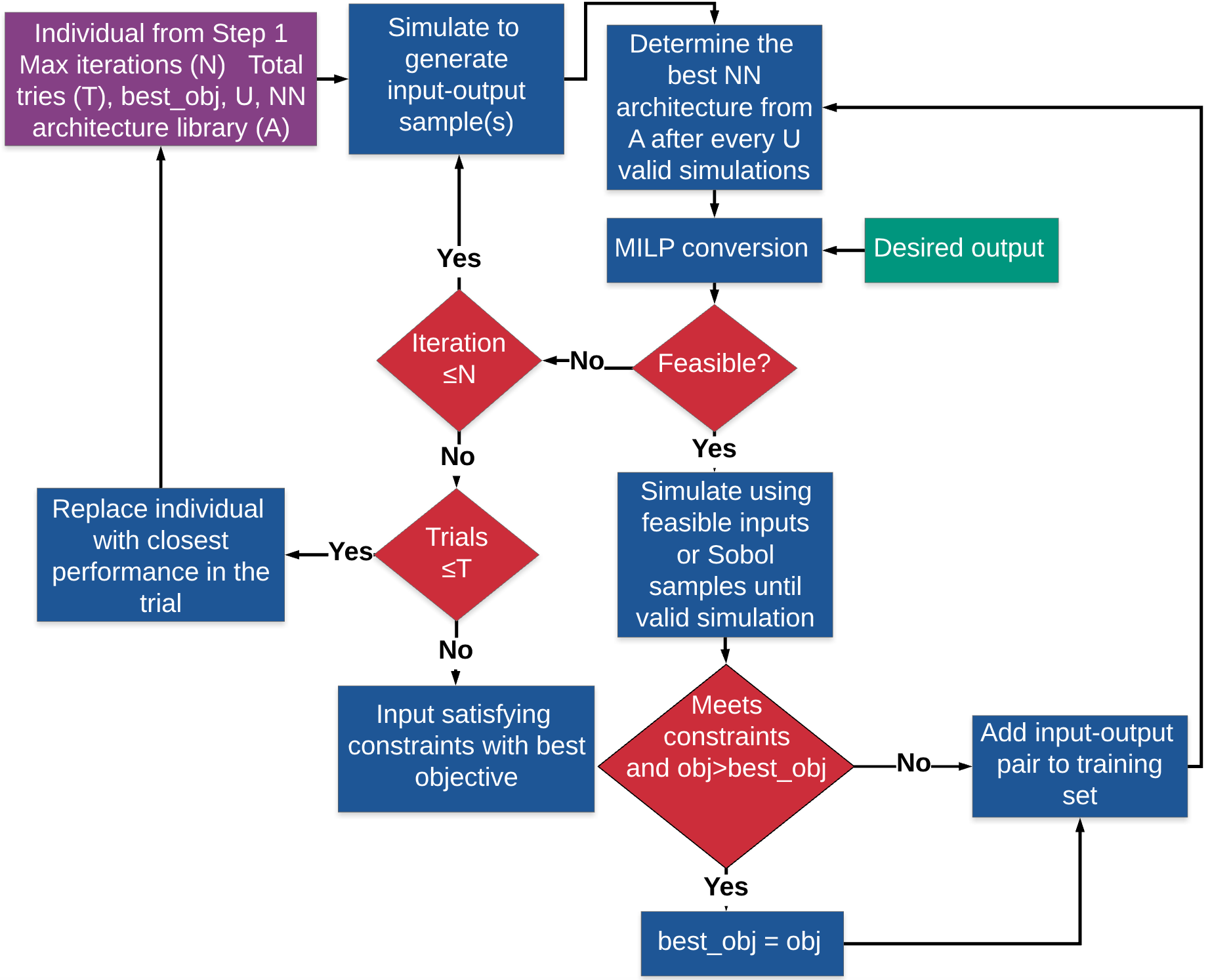}
    \centering
    \caption{Flowchart that illustrates the fine-tuning procedure in Step 2 of ASSENT.}
\label{fig:CNMAFlowchart}
\end{figure}

Fig.~\ref{fig:CNMAFlowchart} shows the procedure to fine-tune the design that Step 1 yields. 
In the flowchart, \textit{N} denotes the number of simulations in a particular trial out of a total 
of \textit{T} trials. We effectively repeat the fine-tuning step \textit{T} times, allowing a maximum 
of \textit{N} valid simulations in a trial. We consider a simulation to be invalid if the output from the simulator yields an error. We represent the best value of the objective attained 
from Step 1 by \textit{best\_obj}.  Next, we generate \textit{Sobol} samples from the range of each 
component in the design from Step 1, followed by simulation of these points to determine their 
outputs. We generate samples around the nominal values of each component (e.g., $\pm 70\%$ of 10 
$\Omega$ for resistor). We clip the samples to ensure that the component values  are within the 
permissible range.  We model the response of the system to the inputs using an NN. We either 
pre-train the NN using the simulations saved from Step 1 and the \textit{Sobol} samples or only on 
the \textit{Sobol}  samples generated around the best design determined from Step 1. We use the 
simulations saved from Step 1 to attain better value of the  objective function in fewer samples. 
However, this is not suitable for some systems where the 
number of samples required may be higher to overcome the inductive bias since Step 1 searches over 
the entire design space. We select the best NN from an NN architecture library that yields the least 
MSE on the validation set.  We update the NN architecture after every \textit{U} valid simulations. 
We convert the NN into an MILP using Eq.~(\ref{eq:MILPConversion}) to see if there is a feasible 
solution.  We determine the constraints from the \textit{desired output} of the system.  The range of 
available components determines the constraints on the input.
We use Gurobi \cite{gurobi} to find a feasible solution to the MILP problem.  If such a solution 
exists, it indicates that the desired output is reachable by the NN from this input. We simulate the 
system with the suggested input. 
If the simulation is invalid, we use Sobol samples until attaining a valid simulation output. 
We update \textit{best\_obj} when the value of the objective 
function improves.  We then add the input-output pair corresponding to the feasible solution to the 
training set. If the solution to the MILP problem is infeasible, we generate a \textit{Sobol} sample 
and add the corresponding output to the training set. The procedure continues until a maximum number 
of permissible simulations (\textit{N}) is exhausted. 

After finishing a trial, the input corresponding to the least absolute sum in terms of fractional 
deviation of the output from the requirement replaces the design from Step 1. Fractional deviation 
is computed as follows:
\begin{equation}
\label{eq: MinDeviationCNMA}
    \begin{aligned}
    \sum_{i}\frac{\left|{\rm obs}_i\ -\ {\rm spec}_i\ \right|}{{\rm spec}_i}\ \mathds{1}_{{\rm \{obs_{i} \lessgtr spec_{i}\}}},
    \end{aligned}
\end{equation}

\noindent
where ${\rm spec}_i$ is the specification for the $i^{th}$ objective or constraint, ${\rm obs}_{i}$ is 
the observed response of the system for the $i^{th}$ objective or constraint.  
$\mathds{1}_{{\rm \{obs_{i} \lessgtr spec_{i}\}}}$ is an indicator function that takes the value 1 in 
the case of violation of ${\rm spec}_i$ determined by ${\rm obs}_i$ (can be greater or less than) and 
0 otherwise. After the last trial, we return the input that corresponds to the output that satisfies 
all the hard constraints and yields the best value of the objective function. 

\vspace{-.2cm}
\section{Experimental Results}

\label{sect:results}
In this section, we first illustrate how ASSENT can be used for component selection in the
design of the following benchmarks: Lunar lander \cite{1606.01540}, Acrobot \cite{1606.01540},
Polak3 \cite{70250}, Waveglider \cite{kraus2012wave},  
Rover path planning \cite{wang2018batched}, and sensor placement for a power grid \cite{openDSS}. Then we show how ASSENT can be used to synthesize
electrical circuits in two cases. In the first case, we use ASSENT for architecture search and
component selection to design a circuit that achieves the functionality of a two-stage
transimpedance amplifier. In the second case, we select the component values for a fixed
architecture for two-stage and three-stage transimpedance amplifiers. 
We implement ASSENT using Keras \cite{chollet2015keras},
Scikit-learn \cite{scikit-learn}, Gurobi \cite{gurobi}, and PyGMO \cite{pygm}. The simulations
are performed on an Intel Xeon processor with 64 GB of DRAM.

\vspace{-.3cm}
\subsection{Component selection for system design}
\label{subsec: ComFixedArch}
We illustrate the selection of component values on various benchmarks and compare their performance 
to that of designs obtained using  CNMA \cite{CNMA2} with 1, 5, and 10 solvers. 

\noindent
\underline{\textit{\textbf{Lunar lander}}}: 
The goal is to maximize the reward obtained by the landing position subject to 
constraints on the fuel used and the time taken to land.  We run Step 1, i.e., GA  for component 
selection with 100 individuals for a maximum of 400 generations. We set the crossover probability 
(\textit{cross}) to 0.9 and mutation rate (\textit{mut}) to 0.1 
that  are typical values for GA. We terminate Step 1 if any of the following stopping criteria are 
met: (1) reward is above 500, (2) after the $10^{th}$ generation, the objective value for reward does 
not improve for 10 generations, or (3) we reach the maximum number of generations. We convert the 
constraints into objectives in Step 1 and formulate the problem as an MOO. There are three objectives 
in Step 1 described next.

\begin{itemize}
    \item Objective for fuel consumption:
    \noindent
\begin{equation}
    \label{eq: fuelObj}
    \ (\frac{F_{\rm{ meas}}}{F_{\rm {max}}} +  \alpha\frac{F_{\rm {meas}}\ -\ F_{\rm {max}}\ }{F_{\rm {max}}})\mathds{1}_{\{F_{\rm {meas}}>F_{\rm max}\}},
\end{equation}
\noindent
where $\alpha$ is set to 15 and represents the penalty for fuel consumption ($F_{\rm meas}$) when it 
exceeds the maximum permissible value ($F_{\rm max}$),  which is 100 units in our case. 
$\mathds{1}_{\{F_{\rm {meas}}>F_{\rm {max}}\}}$ is an indicator function that takes a value of 1 when  
$F_{\rm {meas}}>F_{\rm {max}}$, and $0$ otherwise. 

\item Objective for time taken to land is defined in the same way as fuel consumption. The maximum 
time allowed for landing is 10 units.

\item  Objective for reward is defined as:
\noindent
\begin{equation}
    \label{eq: rewardObj}
    \begin{aligned}
     \frac{400 }{\rm  reward_{attained}} \mathds{1}_{\rm  reward_{attained} > 0} + \\ (\rm abs(reward_{ attained}) +  1000)\mathds{1}_{\rm reward_{attained} \leq 0},
     \end{aligned} 
\end{equation}

\noindent
where $\rm reward_{attained}$ is the reward for a particular choice of component values. As we need 
to maximize the reward, the lower the value of Eq.~(\ref{eq: rewardObj}), the better is the design. 
The first term in Eq.~(\ref{eq: rewardObj}) yields lower value of the objective for a higher reward. 
We choose 400 as the numerator since it is close to the reward attained using sequential CNMA (437.1). The 
second term corresponds to the case for negative reward. We set this term to a large number to 
penalize designs with a negative reward.
\end{itemize}

\noindent
In case of simulation that does not lead to success, we set the objective values to a very large 
number to indicate this fact.

Table \ref{tab:LunarResults} shows the rewards and constraints for the designs obtained using CNMA and 
ASSENT. In the table, CNMA-1, CNMA-5, and CNMA-10 refer to CNMA runs based on 1, 5, and 10
solvers, respectively \cite{CNMA2}. Step 1 requires less time compared to CNMA, since GA does not
need to solve the time-consuming MILP formulation after simulation. 

\begin{table}[h]
\captionsetup{font=footnotesize}
\centering
\caption{Comparison of reward using CNMA and ASSENT for Lunar lander.}
\label{tab:LunarResults}
\resizebox{\columnwidth}{!}{%
\begin{tabular}{|l|L|L|L|L|L|L|}
\hline
Procedure & \# Simulations & Total time & \# Initial samples & fuel $\leq$ 75 & time $\leq$ 10  & reward \\ \hline
CNMA-1      & 2,066  & 1,500s   & 100 & 34.12 & 2.60 & 437.1 \\ \hline
CNMA-5     & 1,358  & 1,500s   & 100 & 59.91 & 2.88 & 469.6 \\ \hline
CNMA-10      & 1,494  & 1,500s   & 100 & 43.35 & 3.92 & 465.1 \\ \hline
Step 1           & 1,400 & 42s    & -  & 52.59 & 2.64 & 451.4 \\ \hline
Step 2           & 583  & 54,000s & 1  & 47.09 & 3.50  & \textbf{475.7} \\ \hline
\end{tabular}
}
\end{table}

In Step 2, we aim to improve the reward further. We set the number of trials ($T$) to 50, maximum 
number of iterations ($N$) to 20 in the first trial, and 50 from the second trial onwards. We terminate Step 2 at the end of 15 hours.
We use all the simulations saved from Step 1 with a reward  greater than 400 to narrow down to 
a promising part of the design space.  We generate Sobol samples within 
$\pm 70$\% of the design attained in Step 1. Since we use the simulation results from Step 1, we 
generate only one additional Sobol sample in each trial. We set the update frequency ($U$) of the NN 
architecture search to 10.  We select an NN with one of the following architectures (A): [(40,20,8), 
(30), (15), (100), (50), (200), (20,20,8), (300), (400), (500), (600)], where the tuples represent 
the number of neurons in the hidden layer(s).  The NN inputs represent component values and the 
outputs represent fuel consumption, time taken to land, and reward.  We use a
\textit{Multi-Layer Perceptron  Regressor} 
from the Scikit-learn \cite{scikit-learn} package along with the Adam optimizer 
\cite{duchi2011adaptive}, an initial learning rate of 0.0001, \textit{adaptive} learning, and a 
\textit{maximum iteration count} of 100,000 to train the NN. We set other parameters to their default 
values. Unless otherwise specified, we use the same setup for all the experiments. 

\begin{figure}[!ht]
    \includegraphics[scale=.30, angle =0]{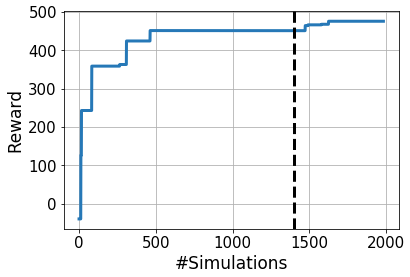}
    \centering
    \caption{Reward vs. number of simulations for Lunar lander in Step 1 (2) to the left (right) of 
the dotted line.}
\label{fig:LLanderPlot}
\end{figure}

The last row of Table \ref{tab:LunarResults} shows the reward attained after fine-tuning.
Step 2 requires more time than 
taken by CNMA, primarily due to the time required for NN architecture search and due to the different 
system configurations used to run the experiments. We improve the value of the reward from 451.4 to 
475.7.
Fig.~\ref{fig:LLanderPlot} plots the highest 
reward attained versus the number of simulations. To the left of the dotted line is the reward 
attained in Step 1 and to the right the reward attained in Step 2.  We observe that fine-tuning helps 
further improve the reward.

\noindent
\underline{\textit{\textbf{Acrobot}}}: 
The objective is to minimize the time required to stabilize ($t_{\rm stabilize}$) 
the system by selecting the design parameters for the two arms of the Acrobot.  We terminate GA if 
$t_{\rm stabilize}$ falls below 3.2s. Other stopping criteria are the same as the ones used in the 
Lunar lander example.  Table \ref{tab:AcrobotResults} shows a comparison of $t_{\rm stabilize}$ 
attained using CNMA and ASSENT.  Step 1 yields a design with $t_{\rm stabilize} = 3.0$s. 

\begin{table}[h]
\centering
\captionsetup{font=footnotesize}
\caption{Comparison of $t_{stabilize}$ using CNMA and ASSENT for Acrobot.}
\label{tab:AcrobotResults}
\resizebox{\columnwidth}{!}{%
\begin{tabular}{|l|l|L|L|l|}
\hline
Procedure              & \# Simulations & Total time & \# Initial samples & $t_{stabilize}$ (s) \\ \hline
CNMA-1          & 247            & 1.4 hrs     & 20                 & 3.4              \\ \hline
CNMA-5          & 285            & 1.4 hrs     & 20                 & 3.2              \\ \hline
CNMA-10          & 520            & 1.4 hrs     & 20                 & 2.8              \\ \hline
Step 1                 & 1,386           & 2.8 hrs    & -                  & 3.0                \\ \hline
Step 2                 & 1,359            & 15 hrs    & 1                  & 2.6              \\ \hline
Step 2- tight bound & 1,293           & 15 hrs  & 1                  & \textbf{2.4}     \\ \hline
\end{tabular}
}
\end{table}

\begin{figure}[!ht]
    \includegraphics[scale=.323, angle =0]{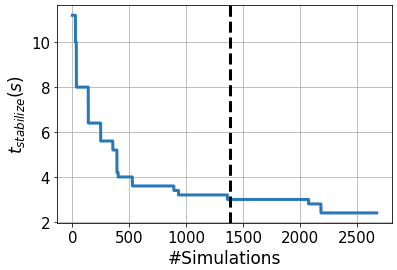}
    \centering
    \caption{$t_{stabilize}$ vs. number of simulations for Acrobot in Step 1 (2) to the left
(right) of the dotted line.}
\label{fig:AcrobotPlot}
\end{figure}

In Step 2, we fine-tune the component values to further reduce $t_{\rm stabilize}$. We use all the 
simulations saved in Step 1 with $t_{\rm stabilize} \leq 5.0$s. We select an NN with one of the 
following architectures (A): [(10), (15), (20), (30), (100)]. 
We reduce the value of  $t_{stabilize}$ to 2.6s.
We further 
explore if tightening the bounds within which we generate the inputs to $\pm 20$\% from 
$\pm 70$\% of the nominal design from Step 1 can reduce $t_{\rm stabilize}$. These results are shown in the last row of 
Table \ref{tab:AcrobotResults}. The modified setup yields a design with $t_{stabilize}$ equal to 2.4s. 
Fig.~\ref{fig:AcrobotPlot} plots the best value of $t_{stabilize}$ versus the 
number of simulations. 

\noindent
\underline{\textit{\textbf{Waveglider}}}: 
The objective in this benchmark is to maximize $V \rm{Boat}_{x}$ subject to the constraints 
$\left|f {\rm{Boat}_x}- f \rm{Glider}_x \right|\le0.05\ast\ f \rm{Boat}_x\land\ V \rm{Boat}_x\geq\ 
V \rm{Boat}_y$. Here, $f \rm{Boat}_x$ ($f \rm{Glider}_x$) is the force of the boat (glider) and
$V \rm{Boat}_x$ ($V \rm{Boat}_y$) is the velocity of the boat in the $x$ ($y$) direction. The 
component values represent the dimensions of the boat and hydrofoil. We define the following 
objectives for GA:

\begin{itemize}
    \item Objective corresponding to the deviation between $f \rm{Boat}_{x}$ and  $f \rm{Glider}_{x}$:
    \noindent
\begin{equation}
    \label{eq: con1Obj}
     (con_{1} +  \alpha\frac{con_{1} - 0.05 }{0.05})\mathds{1}_{con_{1} \geq 0.05},
\end{equation}

\noindent
where   
$con_{1} = \frac{f {\rm{Boat}_{x}} - f {\rm{Glider}_{x}}}{f {\rm{Boat}_{x}} + 1e^{-12}}$, and  
$\alpha = 15$.

\item Objective corresponding to  $V \rm{Boat}_x\geq\ V \rm{Boat}_y$ :
\begin{equation}
    \label{eq: con2Obj}
    \ (con_{2} -  \alpha con_{2})\mathds{1}_{con_{2} \leq 0},
\end{equation}
where  
$con_{2} = V \rm{Boat}_{x} - V \rm{Boat}_{y} $. 

\item The third objective is $-V \rm{Boat}_{x}$. We reverse the sign as we need to maximize the 
velocity of the boat in the $x$ direction. 
\end{itemize}

\begin{table}[h]
\captionsetup{font=footnotesize}
\centering
\caption{Comparison of $V \rm Boat_{x}$ using CNMA and ASSENT for Waveglider.}
\label{tab:Waveglider}
\resizebox{\columnwidth}{!}{%
\begin{tabular}{|l|l|l|L|l|l|l|l|}
\hline
Procedure & \# Simulations & Total time & \# Initial samples &  $V \rm{Boat{_x}}$       & $V \rm{Boat{_y}}$ & $f \rm{Glider{_x}}$\\ \hline
CNMA-1     & 1,920            & 0.69 hr       & 30                 &  \textbf{3.90}          & 2.286    & 107.53     \\ \hline
CNMA-5    & 3,170            & 0.69 hr       & 30                 &  \textbf{3.90}          & 2.286    & 107.48     \\ \hline
CNMA-10    & 3,920            & 0.69 hr       & 30                 &  \textbf{3.90}          & 2.286    & 107.48     \\ \hline
Step 1    & 2,145           & 0.37 hr      & -                  &  3.61          & 2.286    & 248.83     \\ \hline
Step 2    & 2,383            & 15 hrs     & 1                  &  \textbf{3.90}  & 2.286    & 108.95    \\ \hline
\end{tabular}
}
\end{table}

Table \ref{tab:Waveglider} shows the value of $V \rm Boat{_x}$ obtained using CNMA and ASSENT.
Step 1 terminates after 2,145 
simulations in 0.37 hr due to saturation in performance.  

\begin{figure}[!ht]
    \includegraphics[scale=.30, angle =0]{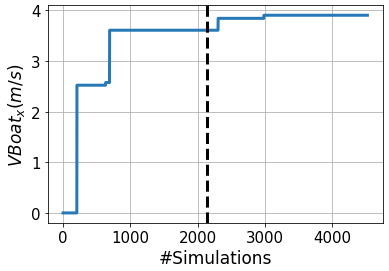}
    \centering
    \caption{The best value of $V \rm{Boat}{_x}$ vs. number of simulations for Waveglider in 
Step 1 (2) to the left (right) of the dotted line.}
\label{fig:WavegliderPlot}
\end{figure}

In Step 2, we use all the valid simulations from Step 1 to train the NN. We select an NN with one of 
the following architectures (A): [(10), (15), (20), (30), (100)]. Besides the three outputs from the simulator, we 
add an additional output when training the NN. This output is an identity mapping of
$f \rm{Boat}_{x}$. It simplifies the MILP formulation by transferring the constraints to the NN 
output. We do not range-normalize the outputs to keep $f \rm{Boat}_{x}$ and $f \rm{Glider}_{x}$ to 
the same scale.  We also increase the learning rate to 0.1 to speed up training. The design 
synthesized at the end of Step 2 meets all the constraints and improves the value of 
$V \rm{Boat}_{x}$ to 3.90 m/s. 
Fig.~\ref{fig:WavegliderPlot} plots the best value of $V \rm{Boat_{x}}$ vs. number of simulations.
 
\noindent
\underline{\textit{\textbf{Polak3}}}: 
The objective is to minimize the maximum value of 10 different transcedental functions.
There are 11 inputs that represent component values. 
The objective of GA is to find the maximum value of the transcedental functions.


\begin{table}[h]
 \captionsetup{font=footnotesize}
 \centering
\caption{Comparison of objective using CNMA and ASSENT for Polak3.}
\label{tab:rvdPolakResults}
\resizebox{\columnwidth}{!}{%
\begin{tabular}{|l|l|l|L|l|}
\hline
Procedure & \# Simulations & Total time & \# Initial samples & Objective      \\ \hline
CNMA-1   & 859            & 2,000s       & 20                 & 5.97          \\ \hline
CNMA-5   & 1,970            & 2,000s       & 20                 & 6.06          \\ \hline
CNMA-10   & 2,700            & 2,000s       & 20                 & 5.98          \\ \hline
Step 1    & 37,397          & 72s      & -                  & \textbf{5.94} \\ \hline
\end{tabular}
}
\end{table}

\begin{figure}[!ht]
    \includegraphics[scale=.30, angle =0]{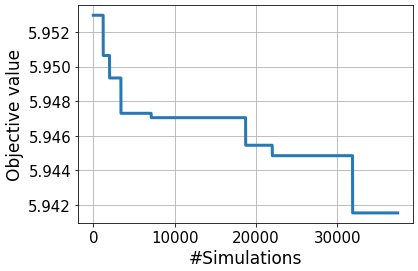}
    \centering
    \caption{The best value of the objective function in Step 1 vs. number of simulations for 
Polak3.}
\label{fig:rvdPolakPlot}
\end{figure}

As simulations are not time-consuming for this problem, we let GA run serially for 400 generations. 
Step 1 requires a total of 37,397 simulations in 72s, as shown in Table \ref{tab:rvdPolakResults}. 
We attain an objective value of 5.94 compared to 5.97 obtained using CNMA. Our solution is closer to the known minimum value of 5.93 for this problem.
Fine-tuning using Step 2 does not improve the value of this objective. This indicates that if the simulations are cheap, Step 1 may alone be sufficient to solve the optimization problem without the need of fine-tuning in Step 2. Fig.~\ref{fig:rvdPolakPlot} 
plots the value of the objective function vs. number of simulations.

\begin{figure}[!ht]
    \includegraphics[scale=.30, angle =0]{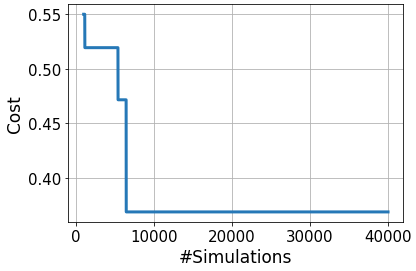}
    \centering
    \caption{Lowest cost vs. number of simulations from the $1000^{th}$ simulation onward in
Step 1 for Rover path planning.}
\label{fig:eboPlot}
\end{figure}

\noindent
\underline{\textit{\textbf{Rover path planning}}}: The objective is to find a 
trajectory that minimizes a cost function by selecting 30 2D points that define the trajectory.  
We use this cost function as the objective for GA. We let GA run for 400 generations since the 
simulations are not time-consuming. Step 1 requires a total of 40,000 simulations in 188s, as shown 
in Table \ref{tab:eboResults}. We obtain an objective value of 0.369 compared to 0.778 obtained using 
CNMA. 
Fine-tuning using Step 2 does not improve the value of this objective further. 
Fig.~\ref{fig:eboPlot} plots the best value of the objective versus the number of simulations.

\begin{table}[h]
 \captionsetup{font=footnotesize}
 \centering
\caption{Comparison of cost function using CNMA and ASSENT for Rover path planning.}
\label{tab:eboResults}
\resizebox{\columnwidth}{!}{%
\begin{tabular}{|l|l|l|L|l|}
\hline
Procedure & \# Simulations & Total time & \# Initial samples & Cost        \\ \hline
CNMA-1     & 1,848              & 9,000s          & 2                  & 1.148                \\ \hline
CNMA-5    & 2,639              & 9,000s          & 2                  & 0.997                \\ \hline
CNMA-10    & 4,454             & 9,000s          & 2                  & 0.778                \\ \hline
Step 1    & 40,000          & 201s    & -                  & \textbf{0.369} \\ \hline
\end{tabular}
}
\end{table}

\noindent
\underline{\textit{\textbf{IEEE 118 bus Sensor placement}}}: The objective is to obtain the 
sensor placement such that sensor readings give the best chance of predicting the line failure 
pattern. The failure pattern is determined by the ``ambiguity" in the sensor pattern. The maximum 
number of sensors is 50. We define the following two objectives for GA: 

\begin{itemize}
    \item The first objective is to capture the constraint on the number of sensors:
\begin{equation}
    \label{eq: con1IEEE118Bus}
  (\rm sensors_{\rm used} -  50)\mathds{1}_{\rm sensors_{\rm used} > 50},    
\end{equation}
where $\rm sensors_{\rm used}$ denotes the number of sensors used in a particular configuration. We apply a penalty of $\alpha = 15$ when using more than 50 sensors.
\end{itemize}

\begin{itemize}
    \item The second objective is to minimize ambiguity as determined by the simulator output.
\end{itemize}

\begin{figure}[!ht]
    \includegraphics[scale=.30, angle =0]{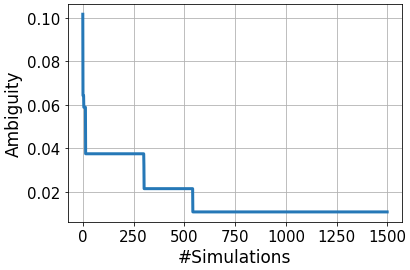}
    \centering
    \caption{The best value of the ambiguity in Step 1 vs. number of simulations for 
the IEEE 118 bus problem.}
\label{fig:ieee118Bus}
\end{figure}

Step 1 terminates after 1,500 simulations due to saturation in performance for 10 generations. Table \ref{tab:IEEE118BusResults} shows that Step 1 achieves an ambiguity of 0.0107 in 1,500 
simulations. The ambiguity is lower by about 50$\%$ than the ambiguity attained using CNMA. 
Fig.~{\ref{fig:ieee118Bus}} plots the value of ambiguity vs. the number of simulations.

\begin{table}[h]
 \captionsetup{font=footnotesize}
 \centering
\caption{Comparison of ambiguity using CNMA and ASSENT for the IEEE 118 bus problem.}
\label{tab:IEEE118BusResults}
\resizebox{\columnwidth}{!}{%
\begin{tabular}{|l|l|l|L|l|}
\hline
Procedure & \# Simulations & Total time & \# Initial samples & Ambiguity       \\ \hline
CNMA-1      & 251              & 1,500s          & 30               &  0.04813                \\ \hline
CNMA-5      & 490              & 1,500s          & 30                  & 0.02139               \\ \hline
CNMA-10      & 380              & 1,500s          & 30                  &  0.02139                \\ \hline
Step 1    & 1,500           & 1,058s      & -                  & \textbf{0.01070} \\ \hline
\end{tabular}
}
\end{table}

\vspace{-.4cm}
\subsection{Circuit design}
In this section, we evaluate how ASSENT performs architecture search and component selection for 
electrical circuits.  We compare our results with those in \cite{wang2018learning} that are 
synthesized by humans, RL, and Bayesian optimization. The technology files that specify the device 
physics for simulation are from \cite{NIPS2019_8519} and are the same as in \cite{wang2018learning}.

\subsubsection{\textbf{Architecture search and component selection}}
\label{subsec:ArchExample} 
\begin{figure}[h]
    \includegraphics[scale=0.28]{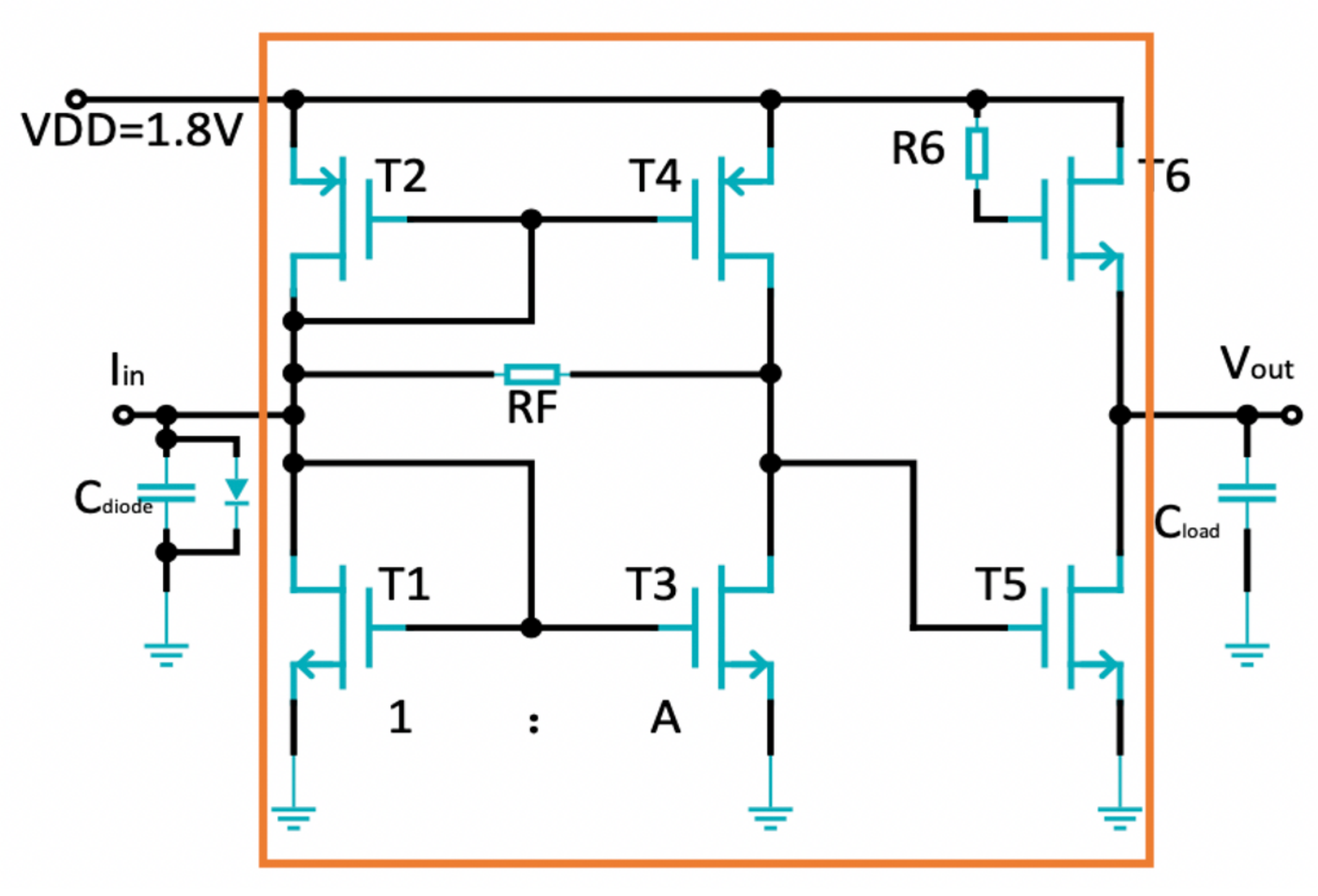}
    \centering
        \caption{Two-stage transimpedance amplifier from \cite{wang2018learning}. Component values 
are selected for devices inside the orange box and the values for those outside are fixed.}
\label{fig:twoStageSeed}
\end{figure}

We use the standard design of a two-stage transimpedance amplifier presented in 
\cite{wang2018learning} as the seed design for our methodology  to take advantage of prior 
human knowledge.  Fig.~\ref{fig:twoStageSeed} shows this seed design. The objectives are to maximize 
the amplifier's bandwidth and minimize the sum of MOSFET gate areas in the circuit while satisfying 
hard constraints on \textit{noise, gain, peaking}, and \textit{power}. The design space comprises 
three components: two types of MOSFETs (PMOS, NMOS) and resistor. In the human-designed circuit, the 
MOSFETs are of minimum length ($0.18\ \mu \rm m$), based on the technology used, whereas the width is 
variable. Our search also uses minimum-length MOSFETs and only selects their width.  We
discretize the design space during architecture search (Step 1) by generating 100 Sobol samples for 
resistors in the $[50, 5\rm k]\ \Omega$ range and width in the $[0.18, 80]\ \mu \rm m$ range. 
In the human-designed circuit, the resistor values are $420~\Omega$ and $3~\rm k\Omega$, and
the width is in the $[0.9, 51]\ \mu \rm m$ range. 
During GA evolution, we allow the circuit to have a 
maximum of 11 nodes and a total of 10 components, whereas the seed design has six nodes and eight 
components, to enable search for novel designs. Since we need to represent 10 components in a 
chromosome, we initialize two additional genes as resistors with a small resistance of 
2.2 $\rm \mu \Omega$ whose terminals are shorted and not connected to any of the nodes in the seed design. This ensures the response of the seed design remains unaffected.

We use GA to evolve a generation of 100 individuals and set the maximum number of generations to 200. 
We choose these numbers to achieve at least about $50\%$ sample efficiency (i.e., $50\%$ fewer 
simulations) compared to the design methodology in \cite{wang2018learning}. We simultaneously search 
for the architecture and the component values, whereas the method in \cite{wang2018learning} only 
searches for component values. Hence, our method searches through a much larger design space, albeit 
with the seed design. We formulate an MOO problem with three objectives: bandwidth, noise, and power.  
The objectives are described next.
\begin{enumerate}
    \item \textit{Bandwidth}: This objective corresponds to the weighted sum of the absolute 
difference between the desired and observed responses, with a passband weight of 40 and a
stopband weight of 1.  We use the following reward/penalty. (a) Assessing the level of reward/penalty is based on the operating region of each MOSFET: 
reward for MOSFET operating in the saturation region, else a penalty. We use a reward of 1 for operation in the saturation region. We apply a penalty of 2 (3) for  operation in the linear (cutoff) region. We sum the rewards and penalty for each MOSFET and divide by the number of MOSFETs. (b) A penalty of 15 is assessed based on the fractional deviation in gain below 
58.1 dB (since 58.1 dB is the gain achieved by RL based synthesis in 
\cite{wang2018learning}). (c) A penalty of 15 is assessed based on the fractional deviation in peaking above 0.963 
(achieved by RL in \cite{wang2018learning}). (d) A penalty of 15 is assessed based on the fractional deviation in bandwidth below 
5.81 GHz (this is slightly higher than the bandwidth achieved by RL in
\cite{wang2018learning}: 5.78 GHz).
    We scale the bandwidth objective by dividing it by the objective of the seed design.
    \item \textit{Noise}: This objective corresponds to the ratio of measured noise and the noise 
achieved by the RL-based design in \cite{wang2018learning} (19.2 $\rm pA/\sqrt{\rm Hz}$). An 
additional penalty of 15 is assessed based on the fractional deviation in noise above this value.
    \item \textit{Power}: This objective corresponds to the ratio of measured power and the power 
achieved by the RL-based design in \cite{wang2018learning} (3.18 $\rm mW$). No penalty is 
assessed in this case due to the large room for optimization available for power consumption.
\end{enumerate}

\noindent
The objective for noise is given by
\begin{equation}
    \label{eq: NoiseObj}
    \ \frac{N_m}{N_{ref}}+\alpha\frac{N_m\ -\ N_{ref}\ }{N_{ref}}\mathds{1}_{\{N_m>N_{ref}\}},
\end{equation}
\noindent
where $N_{ref}$ is the noise of RL-designed circuit, $N_{m}$ is the measured noise, and $\alpha$ = 15 is the penalty. 
Other objectives are defined analogously.

We club together all the sub-objectives related to bandwidth (gain, desired bandwidth, peaking) into 
one to minimize the number of objectives that need to be tackled while capturing the entire frequency 
response. The penalty terms encourage the target response to be better than the response of the 
designs in \cite{wang2018learning}.  In case the simulation is unsuccessful, we set
the objective values to a very large number to indicate this fact.

We use a tournament size of 10, mutation rate of 0.1, and crossover probability of 0.9, same as before.
The evolution stops when one the following criteria are met. (1) The scaled bandwidth objective falls below 0.9 since the aim is to improve it by about $10\%$ relative to that of the seed design. 
(2) The number of generations exceeds 100 and the bandwidth objective stays the same for more 
than 100 generations, indicating saturation in GA performance.
(3) The maximum number of generations (200) is reached.

\begin{figure}[ht]
\includegraphics[scale=0.2, angle = 0]{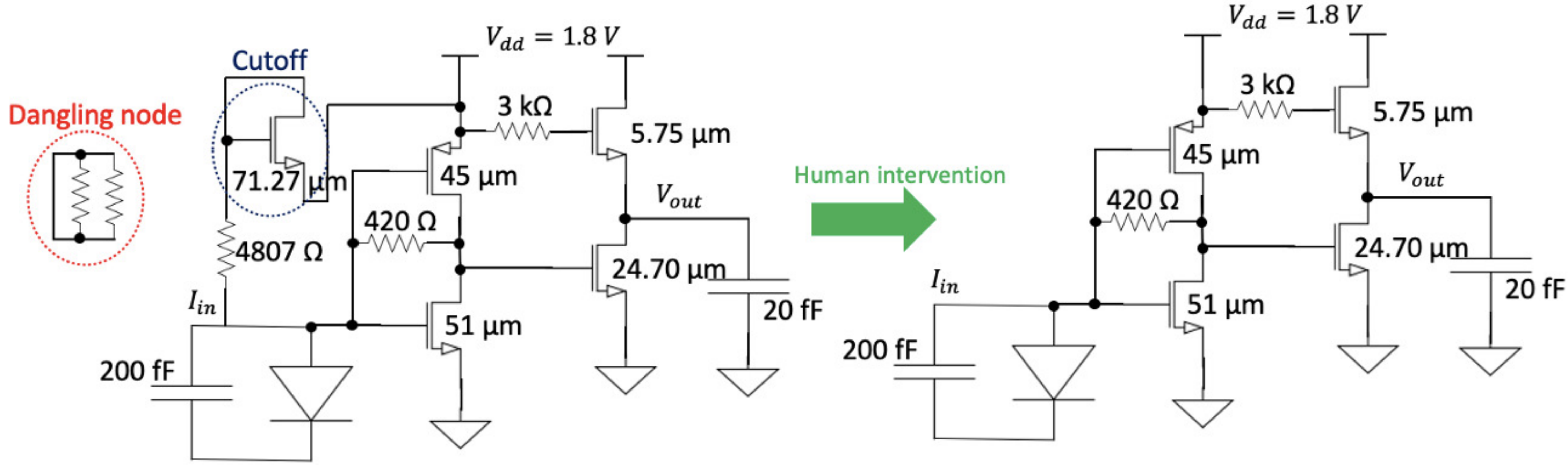}
\centering
\caption{Circuit synthesized by GA followed by human intervention. All MOSFETs are of minimum length 
(0.18 $\mu m$).}
\label{fig:GACircuit}
\end{figure}

We choose the circuit with the best value of the objective for bandwidth as the coarse design in 
Step 1 since this 
circuit addresses several sub-objectives.  We fine-tune the coarse design to meet the specifications 
in the next step. We remove dangling nodes through human intervention as they are redundant. We also 
remove a MOSFET operating in the cutoff region as it does not contribute to gain. 
Fig.~\ref{fig:GACircuit} shows the circuit before and after human intervention. Table 
\ref{tab:GAArchComparisonPaper} shows the values of objectives and constraints before and after human 
intervention. The row named ASSENT Step 1 shows the results before human intervention. GA requires 
less than an hour for synthesis. The second last row shows the results after human intervention.
After the human intervention, the circuit does not meet the hard constraint for peaking. We remedy the 
constraint violation in Step 2.  The seed design has six MOSFETs. This one uses only four MOSFETs.

In Step 2, we fine-tune the component values to maximize the bandwidth while satisfying all the hard 
constraints. We use a slightly modified setup for Step 2 than in Section \ref{subsec: ComFixedArch}. 
Rather than using all the simulations saved from Step 1, we initialize the NN using only Sobol 
samples around the design from Step 1. We use 20 Sobol samples for initialization in the first trial and 1 Sobol sample in each subsequent trial.
We empirically found that using the simulations saved from Step 1 yields inferior results.
We attribute this to the inductive bias imposed in Step 1 due to sampling over the entire
design space. Sobol samples around the design synthesized in Step 1 enable focus on the region of 
interest. In all the experiments on electrical circuits, we select an NN with one of the following 
architectures (A): [(10), (40,20,8), (50)]. We let Step 2 run until 50 trials are over with the other 
setup being the same, as described in the previous section. 

\begin{table*}[!htbp]
 \captionsetup{font=footnotesize}
\centering
\caption{Comparison of circuits synthesized with ASSENT and designs in 
\cite{wang2018learning} (hard constraint violations shown encircled).}
\label{tab:GAArchComparisonPaper}
\resizebox{\textwidth}{!}{\begin{tabular}{|c|c|l|c|c|c|c|c|c|}
\hline
 &
  \#Samples &
  Time &
  Noise ($\rm pA/\sqrt{\rm Hz}$) &
  Gain ($\rm dB\  \Omega$) &
  Peaking ($\rm dB$) &
  Power ($\rm mW$) &
  Gate area (${\rm \mu m}^2$) &
  Bandwidth ($\rm GHz$) \\ \hline
Spec.          & -       & -        & $\leq 19.3$ & $\geq 57.6$ & $\leq 1$ & $\leq 18$ & -       & maximize    \\ \hline
Human Design \cite{wang2018learning} & 1,289,618 & months & 18.6         & 57.7         & 0.927    & 8.11      & 23.11 & 5.95 \\ \hline
DDPG  \cite{wang2018learning}        & 50,000   & 30 GPU hrs & 19.2         & 58.1         & 0.963     & 3.18       & -       & 5.78 \\ \hline
Bayesian Opt.\cite{wang2018learning} & 880     & 30 hrs & \enumber{19.6}         & 58.6         & 0.629     & 4.24       & -       & 5.16 \\ \hline
ASSENT Step 1           & 1,600    & 0.15 hr                & 18.0         & 58.0         & 0.913    & 7.97       & 35.59   & 5.93 \\ \hline
ASSENT Step 1+Human     & -    & \multicolumn{1}{c|}{-} & 17.9         & 58.0          & \enumber{1.042}     & 7.97       & 22.76   & 5.96 \\ \hline

{ASSENT Step 1+2} &
  {4239} & 
  {25.4 hrs} &
  {19.3} &
  {57.6} &
  {0.974} &
  {6.74} &
  {20.24} &
 {\textbf{6.18}} \\ \hline

\end{tabular}}
\end{table*}

\begin{figure}[!ht]
    \includegraphics[scale=.30, angle =0]{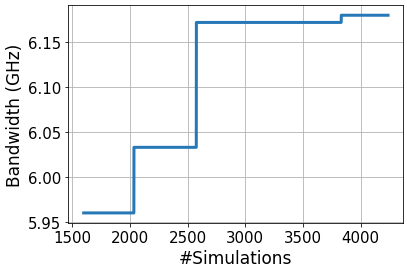}
    \centering
    \caption{The best bandwidth vs. number of simulations in Step 2 when selecting component values
for the architecture obtained using Step 1 for the two-stage transimpedance amplifier.}
\label{fig:cnma2stgArchSearchBest}
\end{figure}

Table \ref{tab:GAArchComparisonPaper} \footnote{Gate area is shown only for designs for which
this information is available. There was a problem in area calculation in \cite{wang2018learning} 
that was confirmed after contacting the authors.} shows a comparison of the design synthesized by 
ASSENT with designs obtained by humans, DDPG, and BO \cite{wang2018learning}. The last row depicts a 
design that satisfies all the hard constraints while maximizing bandwidth. We set the required area 
to less than $0.9\times$ of the human design when maximizing bandwidth.  ASSENT synthesizes a design 
with a bandwidth of 6.18 GHz in 4,239 simulations. It is about  12$\times$ more sample-efficient than 
DDPG and achieves a much higher bandwidth. ASSENT requires a total of 25.4 CPU hours, whereas DDPG 
requires 30 GPU (type not specified) hours \cite{wang2018learning}. 
Fig.~\ref{fig:cnma2stgArchSearchBest} plots the best value of bandwidth vs. the number of 
simulations. We only plot the results for Step 2 to show component selection 
results for a fixed architecture.  




\subsubsection{\textbf{Component selection for fixed architectures}}
\label{subsub: cktCompSelFixArch}
In this section, we use ASSENT to select component values for fixed architectures of a
two-stage and  three-stage transimpedance amplifiers. 

\noindent
\underline{\textit{\textbf{Two-stage transimpedance amplifier}}}: 
We use ASSENT to determine the width of all the MOSFETs and resistors with the goal of 
minimizing the objective functions defined in Section \ref{subsec:ArchExample}.
We discretize the design space in Step 1 by generating 100 Sobol 
samples for resistors in the $[100, 5{\rm k}]\ \Omega$ range and width in the $[0.2, 50]\ \mu \rm m$ 
range.  These value ranges include those for the human-designed circuit. 
Table \ref{tab:GeneForCompSel} shows a chromosome for this case where $W_{i}$ corresponds to
the width of MOSFET $T_{i}$, and $R_{F}$ and $R_{6}$ denote the resistors in 
Fig.~\ref{fig:twoStageSeed}.    

\begin{table}[htb]
 \captionsetup{font=footnotesize}
\caption{Chromosome representation for component selection. The top row shows the parameter name for 
the component and the bottom row the value of that parameter.}
\label{tab:GeneForCompSel}
\resizebox{\columnwidth}{!}{%
\begin{tabular}{|c|c|c|c|c|c|c|c|}
\hline
$W_1$($\rm \mu m$) & $W_2$($\rm \mu m$) & $W_3$($\rm \mu m$) & $W_4$($\rm \mu m$) & $W_5$($\rm \mu m$) & $W_6$($\rm \mu m$) & $R_F$($\rm k\Omega$) & $R_6$($\rm k\Omega$)     \\ \hline
0.39 & 47.86 & 0.59 & 24.71 & 15.18 & 7.40 & 750.8 & 2798.8 \\ \hline
\end{tabular}
}
\end{table}

We evolve a generation of 30 individuals for a maximum of 400 generations. We use a smaller population 
size than in architecture search because the problem is more straightforward due to a smaller 
design space.  The stopping criteria are as follows. (1) The scaled bandwidth with respect to the human design falls below $1$. 
(2) The number of generations exceeds 100 and bandwidth stays the same for more than 100 
generations, indicating saturation.
(3) The maximum number of generations, i.e., 400, is reached.

\begin{figure}[!ht]
    \includegraphics[scale=.30, angle =0]{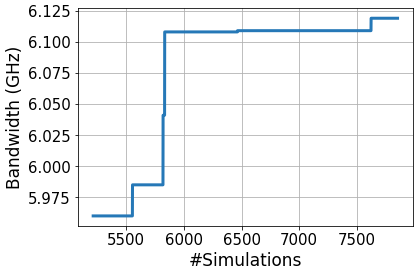}
    \centering
    \caption{Simulation number versus best bandwidth in Step 2 when selecting component values
for the two-stage transimpedance amplifier.}
\label{fig:cnma2stgACompSelBest}
\end{figure}

The second last row of Table \ref{tab:twoStCompSel} shows the value of the objective function 
obtained after Step 1. The bandwidth is inferior to that obtained in human design.  Hence, in Step 2, 
we fine-tune the Step 1 design through component selection to improve its performance. The last row 
of Table \ref{tab:twoStCompSel} shows the design synthesized using ASSENT. ASSENT is more than 
6$\times$ sample-efficient than DDPG and achieves the highest bandwidth among all considered designs. 
Fig.~\ref{fig:cnma2stgACompSelBest} plots the best bandwidth value vs. the number of
simulations used in Step 2.

\begin{table*}[h]
 \captionsetup{font=footnotesize}
\centering
\caption{Comparison of ASSENT designs with those in \cite{wang2018learning} for the two-stage 
transimpedance amplifier (hard constraint violation shown encircled).} 
\label{tab:twoStCompSel}
\resizebox{\textwidth}{!}{\begin{tabular}{|c|c|l|c|c|c|c|c|c|}
\hline
 &
  \#Samples &
  Time &
  Noise ($\rm pA/\sqrt{\rm Hz}$) &
  Gain ($\rm dB\  \Omega$) &
  Peaking ($\rm dB$) &
  Power ($\rm mW$) &
  Gate area (${\rm \mu m}^2$) 
  &
  Bandwidth ($\rm GHz$) \\ \hline
Spec          & -       &       & $\leq 19.3$ & $\geq 57.6$ & $\leq 1$ & $\leq 18$ & -       & maximize    \\ \hline
Human Design \cite{wang2018learning}  & 1,289,618 & months & 18.6         & 57.7         & 0.927    & 8.11      & 23.11 & 5.95 \\ \hline
DDPG \cite{wang2018learning}          & 50,000   & 30 GPU hrs & 19.2         & 58.1         & 0.963     & 3.18       & -       & 5.78 \\ \hline
Bayesian Opt. \cite{wang2018learning} & 880     & 30 hrs & \enumber{19.6}         & 58.6         & 0.629     & 4.24       & -       & 5.16 \\ \hline
ASSENT   Step 1        & 5,214    & 0.7 hr & 19.2         & 58.5         & 0.645    & 5.51      & 17.86   & 5.71 \\ \hline
{ASSENT (Step 1+2)} &
  {7,853} &
  {77 hrs} &
  {19.2} &
  {57.6} &
  {0.949} &
  {5.60} &
  {{18.15}} &
  {\textbf{6.12}} \\ \hline 
\end{tabular}}
\end{table*}

\begin{figure}[!ht]
    \includegraphics[scale=0.20]{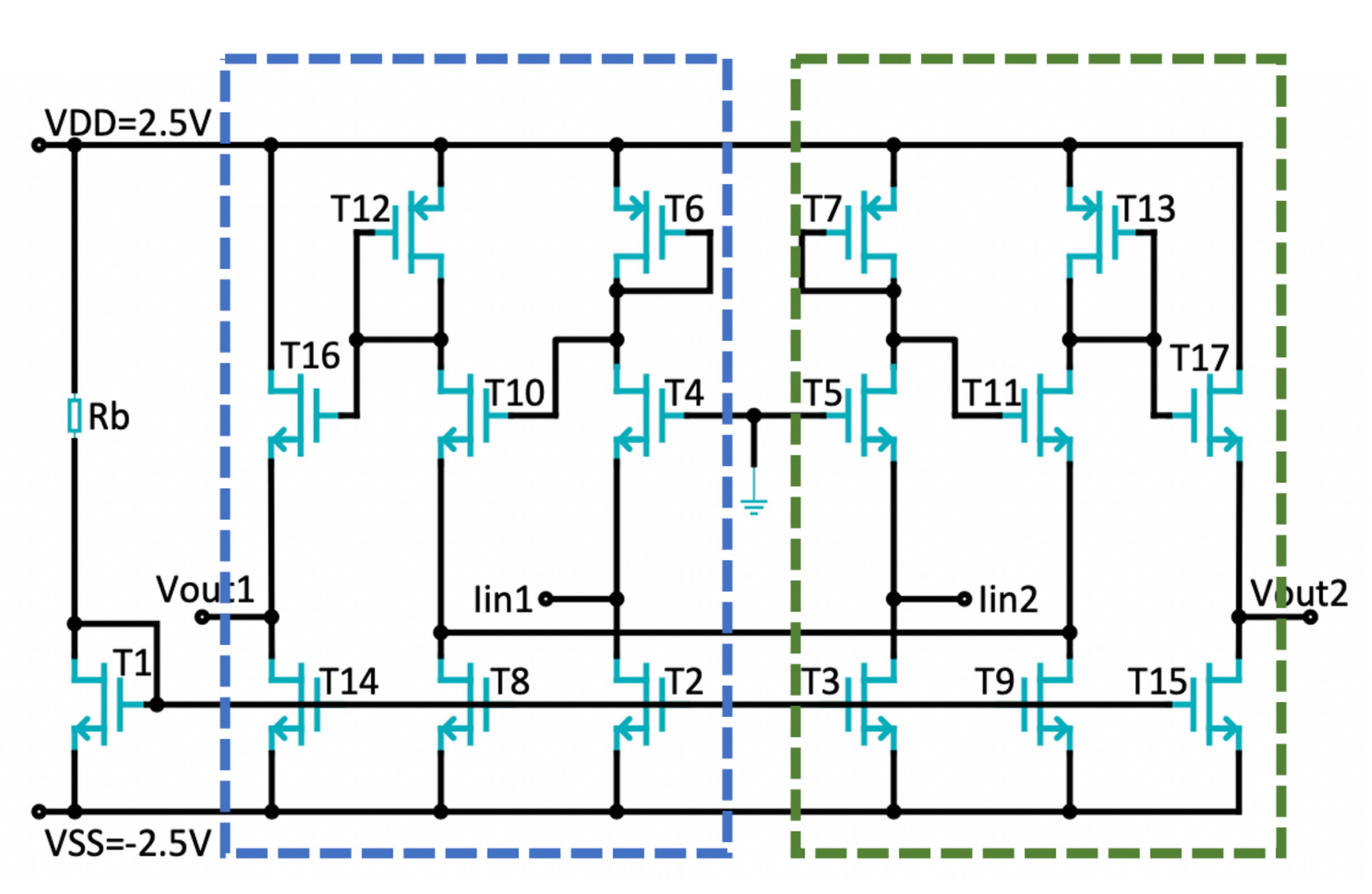}
    \centering
    \caption{Schematic of a three-stage differential transimpedance amplifier 
\cite{wang2018learning}.}
\label{fig:ThreeStageSchematic}
\end{figure}

\noindent
\underline{\textit{\textbf{Three-stage transimpedance amplifier}}}:  We select  the component
values for the  three-stage transimpedance amplifier shown in 
Fig.~\ref{fig:ThreeStageSchematic}, which is adapted from \cite{wang2018learning}.  The blue and green 
dotted boxes contain subcircuits that are mirror images of each other.  Hence, we obtain the
component values for only one subcircuit and mirror them in the other.  We determine the width 
and length of all the MOSFETs, including the bias transistor T1.  There are 19 components in
all: width/length of nine MOSFETs and a resistor \emph{Rb}.

\vspace{-.0cm}
The objective is to minimize the sum of the gate areas of all the MOSFETs while meeting hard 
constraints for \textit{gain}, \textit{bandwidth}, and \textit{power}.  We discretize the
design space during component selection in Step 1 by generating 100 Sobol samples with width in the 
$[2, 30]\ \mu \rm m$, length in the $[1, 2.2]\ \mu \rm m$, and resistor in the 
$[50{\rm k}, 500{\rm k}]\ \Omega$ range. The human-synthesized circuit has width in the 
$[2, 44]\ \mu \rm m$ range, length in the $[1, 2]\ \mu \rm m$ range, and a 
resistor of $291\ {\rm k}\Omega$.  We reduce the design space for width by around $30\%$ to encourage 
Step 1 to obtain designs with a smaller area. We also round the length  to the nearest $0.2$ units 
since the technology only permits the length to be an integer multiple of $0.2\ \mu \rm m$. We use 
this rounding in Step 2 as well.

We have the following objectives:

\begin{enumerate}
    \item \textit{Bandwidth}: We set the passband to $90~{\rm MHz}$ with a target gain of 
$35.57\  {\rm k}\Omega$ (this corresponds to a gain of $85\  {\rm dB}$ for the half-circuit whereas the minimum 
gain required for its valid design is $80\  {\rm dB}$ or $20\  {\rm k}\Omega$ differential
gain). We use the following rewards/penalties. (a) A reward for a MOSFET operating in the
saturation region, else a penalty is applied. We use a reward of 1 for operation in the saturation 
region. We apply a penalty of 5 (7) for operation in the linear (cutoff) region. Since the number of 
MOSFETs is larger than in the earlier design, we slightly increase these numbers to encourage MOSFET 
operation in the saturation region. (b) Penalty of 15 based on fractional deviation in gain below 
80 dB ($20\ {\rm k\Omega}$) for the half circuit in Fig.~\ref{fig:ThreeStageSchematic}. 
(c) Penalty of 15 based on fractional deviation in bandwidth below $90~ {\rm MHz}$.
    \item \textit{Area}: We define the area objective as the ratio of the sum of the areas of the MOSFET in one 
of the mirrored regions and the area of the bias MOSFET (T1) and the corresponding sum of areas in the 
human-synthesized circuit. 
    \item \textit{Power}: We define this objective as the ratio between measured power and the power 
consumed by the human-designed circuit, which is $1.37~\rm mW$.  We levy a penalty of 15 based on the 
fractional deviation in power above this value. 
\end{enumerate}

\begin{figure}[!ht]
    \includegraphics[scale=.30, angle =0]{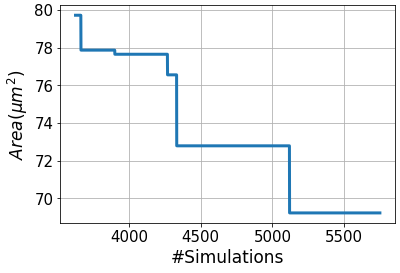}
    \centering
    \caption{The best area vs. number of simulations in Step 2 when selecting component values
for the three-stage transimpedance amplifier.  Results are shown when the area drops below 
80 $\mu m^{2}$.}
\label{fig:cnma3stgBestArea}
\end{figure}

The second last row of Table \ref{tab:ThreeStageCompSelCNMA} shows results of the circuit synthesized 
using Step 1. This circuit violates the hard constraint on power and is hence not an acceptable 
design. The area of this circuit is also much higher than that of the human-designed circuit.
Hence, in the next step, we fine-tune the component values to minimize the area while satisfying all 
the hard constraints. The last row of Table \ref{tab:ThreeStageCompSelCNMA} shows the results
obtained by ASSENT. We set the  initial requirement for the area to $80\ \rm {\mu m}^2$. ASSENT 
synthesizes a design which satisfies all the hard constraints with an area that is around $23\%$
lower than the design obtained using DDPG. ASSENT is around $7\times$ more sample-efficient than 
DDPG. Fig.~\ref{fig:cnma3stgBestArea} plots the best area value vs. the number of simulations. The area is computed after rounding MOSFET length to the nearest 0.2 $\mu$m.

\begin{table*}[]
 \captionsetup{font=footnotesize}
\caption{Comparison of designs synthesized using ASSENT with those synthesized
in \cite{wang2018learning} for the three-stage transimpedance amplifier (hard constraint
violations shown encircled).} 

\label{tab:ThreeStageCompSelCNMA}
\centering
\begin{tabular}{|c|c|c|c|c|c|c|}
\hline
\multicolumn{1}{|c|}{} &
  \multicolumn{1}{c|}{\#Samples} &
  Time &
  \multicolumn{1}{c|}{Bandwidth ($\rm MHz$)} &
  \multicolumn{1}{c|}{Gain ($\rm k \Omega$)} &
  \multicolumn{1}{c|}{Power ($\rm mW$)} &
  \multicolumn{1}{c|}{Gate area (${\rm \mu m}^2$)} \\ \hline
\multicolumn{1}{|c|}{Spec} &
  \multicolumn{1}{c|}{-} &
  - &
  \multicolumn{1}{c|}{$\geq 90$} &
  \multicolumn{1}{c|}{$\geq 20$} &
  \multicolumn{1}{c|}{$\leq 3$} &
  \multicolumn{1}{c|}{-} \\ \hline
\multicolumn{1}{|c|}{Human Design \cite{wang2018learning}} &
  \multicolumn{1}{c|}{10,000,000} &
  months &
  \multicolumn{1}{c|}{90.1} &
  \multicolumn{1}{c|}{20.2} &
  \multicolumn{1}{c|}{1.37} &
  \multicolumn{1}{c|}{211.0} \\ \hline
DDPG \cite{wang2018learning} &
  40,000 &
  40 GPU hrs &
  92.5 &
  20.7 &
  2.50 &
  90.0 \\ \hline
Bayesian Opt. \cite{wang2018learning} &
  1,160 &
  40 hrs &
  \enumber{72.5} &
  21.1 &
  \enumber{4.25} &
  130.0 \\ \hline
{ASSENT Step 1} & {3,060} & {0.5 hr} & {90.2}    & {22.9}               & {\enumber{4.96}} & {337.0}     \\ \hline  
{ASSENT (Step 1+2)} &
  {5,749} &
  {84.7 hrs} &
  {90.5} &
  {20.1} &
  {3.00} &
  {\textbf{69.2}} \\ \hline
\end{tabular}
\end{table*}

\vspace{-.075cm}
\section{Conclusion}
\label{sec:conclusion}
In this article, we formulated nonlinear system design as an MOO problem and proposed a framework called ASSENT to solve it. We used gradient-free search through GA for exploration and CNMA for sample efficiency. CNMA achieves sample efficiency by connecting neural networks with MILP solvers in a new learning-from-failure feedback loop.
The methodology provides a flexible framework for discovering efficient 
architectures and component values or simply the component values for a fixed architecture. Using 
this framework, we achieve the same or 
improve the value of the objective functions compared to designs synthesized using 
CNMA by up to 53$\%$. We also synthesized electrical circuits with the best value of the objective 
function while improving sampling efficiency by 6-12$\times$.  As part of future work, we plan to 
formulate nonlinear system design as graph problems in order to explore different architectures 
by manipulating these graphs and other sample-efficient techniques for \emph{active learning}.  We 
also plan to enhance the efficacy of the methodology by combining it with RL.

\vspace*{2mm}
\noindent

{\bf Acknowledgment:}
We would like to thank Hanrui Wang for providing details of his work and the simulation environment 
used in \cite{wang2018learning}. The authors would also  like to acknowledge the help provided by Sanjai Narain, Emily Mak, and Karthik Narayan with CNMA and various benchmarks used for evaluation of ASSENT. The simulations presented in this article were performed on 
computational resources managed and supported by Princeton Research Computing at 
Princeton University.

\bibliographystyle{IEEEtran}

\end{document}